\documentclass[preprint,11pt]{article}

\usepackage[utf8]{inputenc}
\usepackage{a4wide}
\usepackage[fleqn]{amsmath}
\usepackage{amssymb}
\usepackage{amsfonts}
\usepackage{mathtools}
\usepackage{amsthm}
\usepackage{graphicx}
\usepackage{slashed}
\usepackage{multirow}
\usepackage{listings}
\usepackage{color}
\usepackage{cite}

\usepackage{fancyhdr}

\usepackage[section]{placeins}
\usepackage{float}
\usepackage{authblk}
\restylefloat{table}

\usepackage{booktabs,multirow,tabularx}
\usepackage{multirow}
\usepackage{slashed}
\usepackage{rotating}
\usepackage{etex,etoolbox}

\newcolumntype{L}[1]{>{\raggedright\let\newline\\\arraybackslash\hspace{0pt}}m{#1}}
\newcolumntype{C}[1]{>{\centering\let\newline\\\arraybackslash\hspace{0pt}}m{#1}}
\newcolumntype{R}[1]{>{\raggedleft\let\newline\\\arraybackslash\hspace{0pt}}m{#1}}
\newcolumntype{N}{@{}m{0pt}@{}}

\numberwithin{equation}{section}

\usepackage[colorlinks=true,linkcolor=blue,citecolor=blue]{hyperref}
\usepackage[title,titletoc]{appendix}

\definecolor{shaded}{RGB}{245,245,245}
\def\A{\mathcal{A}}

\def\N{\mathcal{N}}
\def\O{\mathcal{O}}

\def\nn{\nonumber \\}

\newcommand{\Ninja}{\textsc{Ninja}}
\newcommand{\CutTools}{{\sc\small CutTools}}
\newcommand{\IREGI}{{\sc\small IREGI}}
\newcommand{\MadLoop}{{\sc\small MadLoop}}
\newcommand{\PJFry}{{\sc\small PJFry++}}
\newcommand{\Golem}{{\sc\small Golem95}}
\newcommand{\COLLIER}{{\sc\small COLLIER}}
\newcommand{\Samurai}{{\sc\small Samurai}}
\newcommand{\ALOHA}{{\sc\small ALOHA}}
\newcommand{\MGaMC}{{\sc\small MG5aMC}}
\newcommand{\MGaMClong}{{\sc\small MadGraph5\_aMC@NLO}}
\newcommand\prompt{{\tt MG5\_aMC>}}
\newcommand\sss{\scriptscriptstyle}
\newcommand\as{\alpha_{\sss S}}
\newcommand\tnum{$t_\textrm{num}$}
\newcommand\tred{$t_\textrm{red}$}

\newcommand{\NA}{{\text{N/A}}}

\chardef\MyArticleWithColor=\pdfcolorstackinit page direct{0 g}

\title{\textbf{Tensor integrand reduction via Laurent expansion}}
\date{}

\author[1]{Valentin Hirschi}
\author[2]{Tiziano Peraro}
\affil[1]{\emph{\small SLAC, National Accelerator Laboratory, %
2575 Sand Hill Road, %
Menlo Park, %
CA 94025-7090, USA}}
\affil[2]{\emph{\small Higgs Centre for Theoretical Physics, %
School of Physics and Astronomy, %
The University of Edinburgh, %
Edinburgh EH9 3JZ, Scotland, UK }}

\begin{document}

\maketitle
\thispagestyle{fancy}

\begin{abstract}
  We introduce a new method for the application of one-loop integrand reduction via the Laurent expansion algorithm, as implemented in the public {\sc C++} library \Ninja. We show how the coefficients of the Laurent expansion can be computed by suitable contractions of the loop numerator tensor with cut-dependent projectors, making it possible to interface \Ninja\ to any one-loop matrix element generator that can provide the components of this tensor.  We implemented this technique in the \Ninja\ library and interfaced it to \MadLoop, which is part of the public \MGaMClong\ framework.  We performed a detailed performance study, comparing against other public reduction tools, namely \CutTools, \Samurai, \IREGI, \PJFry\ and \Golem.  We find that \Ninja\ outperforms traditional integrand reduction in both speed and numerical stability, the latter being on par with that of the tensor integral reduction tool \Golem\ which is however more limited and slower than \Ninja.  We considered many benchmark multi-scale processes of increasing complexity, involving QCD and electro-weak corrections as well as effective non-renormalizable couplings, showing that \Ninja's performance scales well with both the rank and multiplicity of the considered process.
\end{abstract}

\clearpage

\tableofcontents

\section{Introduction}
\label{sec:introduction}

Scattering amplitudes in quantum field theory describe the fundamental
interactions between elementary particles and provide a powerful
 way for inferring theoretical models form high-energy
phenomenology and vice-versa.  At the scales probed by modern colliders, such as the
Large Hadron Collider (LHC) at CERN, scattering amplitudes can be
computed in perturbation theory as a Taylor expansion in the coupling
constants.  Leading-order results are
plagued by very large theoretical uncertainties and as such they are often not reliable enough for
direct comparisons with experimental results.  Phenomenological
studies can significantly benefit from 
theoretical predictions at next-to-leading order accuracy or beyond,
which are however complicated by several factors, a crucial one being
represented by quantum corrections to amplitudes
computed via loop integrals.  The calculation of these integrals can
be extremely challenging, especially for processes involving many
external legs and physical scales.  Such processes can however be of
great relevance, both for testing the Standard Model in unexplored
regions of phase space and for simulating backgrounds to
signals of interesting (new) physics.  This makes
the calculation of loop amplitudes a very active field of research.

A solution to the problem of computing generic one-loop integrals is
offered by \emph{integrand reduction}~\cite{Ossola:2006us,Ellis:2007br,Giele:2008ve}.  

Integrand reduction
methods rewrite one-loop integrands as a linear combination of terms
in an integrand basis, each of which has five or less loop
propagators and yields either a vanishing integral or a known Master
Integral.  The numerical evaluation of these master integrals is possible by
means of public libraries such as
\textsc{OneLOop}~\cite{vanHameren:2010cp},
\textsc{Golem-95}~\cite{Binoth:2008uq,Guillet:2013msa},
\textsc{LoopTools}~\cite{Hahn:1998yk} and
\textsc{QCDLoop}~\cite{Ellis:2007qk} (the last two use the \textsc{FF}
library~\cite{vanOldenborgh:1990yc} internally).  Because the form of
the integrand basis is universal and independent of the process or the
number of external legs, the algorithm can be applied to any one-loop
scattering amplitude in any Quantum Field Theory.
The \emph{coefficients} of this decomposition, also known as
\emph{integrand decomposition} or \emph{OPP
  decomposition}~\cite{Ossola:2006us}, can be efficiently computed by
evaluating the integrand on \emph{multiple-cuts}, i.e.\ values of the
loop momentum such that a subset of the internal loop propagator denominators
vanish.  This algorithm is based on repeated numerical evaluations of
the integrands and the solution of the resulting subsystems of equations for the
coefficients.  The method has been implemented in the public codes \textsc{CutTools}~\cite{Ossola:2007ax} and
\textsc{Samurai}~\cite{Mastrolia:2010nb}, and has been used within
several automated frameworks
\cite{Hahn:1998yk,vanHameren:2009dr,Bevilacqua:2011xh,Berger:2008sj,Cullen:2011ac,Cascioli:2011va,Badger:2010nx,Badger:2012pg,Heinrich:2010ax}
for producing a wide variety of phenomenological results.
\MadLoop~\cite{Hirschi:2011pa}, part of the \MGaMClong~\cite{Alwall:2014hca} framework (abbreviated \MGaMC\ henceforth), is an example of such tool. It automatically generates one-loop matrix elements and computes them using both traditional OPP reduction (\CutTools\ and \Samurai) and tensor integral reduction~\cite{Passarino:1978jh,Davydychev:1991va} (as implemented in the tools \Golem~\cite{Cullen:2010hz}, \PJFry~\cite{Fleischer:2011bi} and \IREGI).
\MadLoop\ features an in-house implementation of the {\sc\small OpenLoops} method~\cite{Cascioli:2011va} using a modified version of the \ALOHA~\cite{deAquino:2011ub} module to compute the components of the tensor integrand numerator.

More recently, a new approach to one-loop integrand reduction has been
developed, namely the \emph{integrand reduction via Laurent expansion}
method~\cite{Mastrolia:2012bu}, which elaborates on techniques first
proposed in~\cite{Forde:2007mi,Badger:2008cm} for analytic
calculations.  Within this approach, the computation of the
coefficients of the Master Integrals is significantly simplified by
performing a Laurent expansion of the integrands with respect to the
components of the loop momentum which are not constrained by the
multiple-cut conditions.  Since loop integrands are rational
functions of the loop components, within semi-numerical calculations
the semi-analytic Laurent expansion can be performed via a simplified
polynomial division algorithm between the expansion of the numerator
and the loop denominators.  Such a technique has been implemented in the
public \textsc{C++} library \textsc{Ninja}~\cite{Peraro:2014cba},
which combined to the one-loop package
\textsc{GoSam}~\cite{Cullen:2011ac,Cullen:2014yla,vanDeurzen:2013saa} has been used for
producing several phenomenological results for complicated processes 
both within the Standard Model and beyond.

The Laurent expansion reduction algorithm implemented in \textsc{Ninja} needs as
input procedures which return the leading terms of the above mentioned parametric Laurent
expansions of the numerator of the integrand.  Generating such input is straightforward and 
easy-to-automate for analytic one-loop generators such as
\textsc{GoSam}~\cite{Cullen:2011ac,Cullen:2014yla} and
\textsc{FormCalc}~\cite{Hahn:1998yk}, but it has so far prevented
other one-loop tools following a more numerical approach from using reduction via Laurent expansions.

However, as already noted in ref.s~\cite{Mastrolia:2012bu,Peraro:2014cba}, the only explicit
analytic information needed by \textsc{Ninja} about the integrand
is its dependence on the loop momentum (and not, for instance,
on the external kinematics or polarization states), which is always known in the case of tensor-based loop generators,
regardless of whether the entries of the tensors are generated
analytically or numerically.

We present an efficient numerical algorithm for
constructing the Laurent expansions needed by \textsc{Ninja} directly from the
(numerical) entries of loop tensor numerators.  In particular, each
term of these expansions can be computed by contracting the tensor
numerator with specific cut-dependent tensorial projectors.  In a numerical implementation,
these projectors can in turn be constructed at run-time by means of
simple recursive formulas, from lower to higher ranks.  This algorithm
has been implemented within the \textsc{Ninja} library and the only
inputs it needs for the reduction, besides the definition of the loop
propagators, are the numerical components of the tensor numerator. 
This allowed to interface \Ninja\ to \MadLoop\ whose ability to compute these components is now sufficient, as demonstrated in this paper.

In sect.~\ref{sec:tensorintegrands}, we recall the definition of the
tensor integrand and in sect.~\ref{sec:semi-numer-integr} we briefly review the integrand reduction
technique via Laurent expansion. We fix the notation and introduce the computational techniques for building symmetric tensors in sect.~\ref{sec:symmetric-tensors} which we use in sect.~\ref{sec:tens-proj-laur} to derive formulas for the projection of the tensor numerator onto the coefficients of the Laurent expansion. Details on the implementation of this projection in \Ninja\ as well as its interface to \MadLoop\ are provided in sect.~\ref{sec:implementation}. We present a detailed study of the stability and timing performances of the combination of \MadLoop\ and \Ninja\ in sect.~\ref{sec:applications} and we give our conclusions in sect.~\ref{sec:conclusions}.
 
\section{Tensor integrands}
\label{sec:tensorintegrands}

A generic one-loop amplitude can be written as a sum of $n$-point
integrals of the form
\begin{equation}
  \label{eq:1}
  \A = \int d^d \bar q \frac{\N(\bar q)}{D_1\cdots D_n}.
\end{equation}
The numerator $\N$ of the integrand is a \emph{polynomial} in the
components of the $d$-dimensional loop momentum $\bar q$, with
$d=4-2\epsilon$, while the denominators $D_i$ correspond to Feynman
loop propagators and they have the general quadratic form
\begin{equation} \label{eq:loopden}
  D_i = (\bar q + p_i) - m_i^2,
\end{equation}
where $p_i$ is a linear combination of external momenta and $m_i$ is
the mass of the particle propagating in the loop (which can be complex-valued when treating unstable particles in the loop within the \emph{complex mass scheme}~\cite{Denner:2006ic,Denner:2005fg}).
One can split the $d$-dimensional loop momentum $\bar q$ into a
four-dimensional part $q$ and a $(-2\epsilon)$-dimensional part
$\vec{\mu}$,
\begin{equation}
  \label{eq:2}
  \bar q = q+\vec{\mu}, \qquad \bar q^2 = q^2-\mu^2.
\end{equation}
The numerator thus becomes a polynomial in the four-dimensional
components of $q$ and $\mu^2$,
\begin{equation}
  \label{eq:4}
  \N(\bar q) = \N(q,\mu^2).
\end{equation}
We define a four-dimensional \emph{tensor numerator} as a numerator
cast into the form
\begin{equation}
  \label{eq:tensornum}
  \N(q) \equiv \N(q,\mu^2)\Big|_{\mu^2=0} = \sum_{r=0}^R \tilde \N^{(r)}_{\mu_1\cdots \mu_r} q^{\mu_1}\cdots q^{\mu_r}
\end{equation}
where the tensor coefficients $\tilde \N^{(r)}_{\mu_1\cdots \mu_r}$
are independent of the loop momentum.  In principle one could consider
the more general definition of a $d$-dimensional tensor numerator,
i.e.\ a linear combination of terms with the form of the r.h.s.\ of
Eq.~\eqref{eq:tensornum} multiplied by powers of $\mu^2$.  Although it
is straightforward to generalize the results of this paper to
$d$-dimensional numerators, we will only consider the four-dimensional
case since tensor-based one-loop matrix element generators typically compute 
the contributions arising from $\mu^2$ terms in the numerators (also known as $R_2$
contributions) separately using special tree-level Feynman
rules~\cite{Draggiotis:2009yb}.

In the definition in Eq.~\eqref{eq:tensornum}, a tensor numerator is
therefore defined by a sum of tensors homogeneous in rank $r$, for
$r=0,\ldots,R$.  The maximum rank $R$ satisfies $R\leq n$ for
renormalizable theories, up to a gauge choice.  In this paper we will
consider a more general case, i.e.\ $R\leq n+1$, allowing for up to one
effective non-renormalizable vertex in the loop.

In a naive implementation, a generic tensor-numerator of rank $R$
would be defined by $\sum_{r=0}^R4^r=(4^{R+1}-1)/3$ entries, growing
exponentially with the rank.  However, since the tensor described by
Eq.~\eqref{eq:tensornum} is completely symmetric by construction, we
can cast the numerator in the alternative form
\begin{equation}
  \label{eq:symtensornum}
  \N(q)=\sum_{r=0}^R\; \sum_{\mu_1\leq \cdots \leq \mu_r}\; \N^{(r)}_{\mu_1\cdots \mu_r} q^{\mu_1}\cdots q^{\mu_r}
\end{equation}
where the total number of $\mu_i$-symmetric coefficients
$\N^{(r)}_{\mu_1\cdots \mu_r}$ for $r=0,\ldots,R$ (only defined for
$\mu_1\leq\mu_2\leq\cdots \leq \mu_r$) is now
${{\sum_{r=0}^R} {r+3 \choose r}}={R+4 \choose R}$, which only grows
polynomially (namely as $R^4$) with the rank.

In order to clarify the notation, we consider the following example of an arbitrarily chosen $d$-dimensional numerator function
\begin{equation}
    \N(\bar{q}) ={}  4\bar{q}^4 + 2\bar{q}\cdot p_1\, \bar{q}\cdot p_2\, \bar{q}^2 + 2p_1\cdot p_2\, \bar{q}^2 + 3\bar{q}\cdot p_1\, \bar{q}\cdot p_2 + m_X^4 
\end{equation}
which can be recast into the tensorial structure
\begin{flalign}
\label{eq:tensornumexample}
  \N(q,\mu^2) 
  ={}& \underbrace{\left\{ 4\, g^{\mu\nu} g^{\rho\sigma}+2\, p_1^\mu\, p_2^\nu\, g^{\rho\sigma} \right \}_{\textrm{sym}}  }_{\N^{(4)}_{\mu\nu\rho\sigma}(\{p_i\})} q^\mu q^\nu q^\rho q^\sigma \nonumber\\
  &+\underbrace{\left\{ (2\, p_1\cdot p_2\,- 8\, \mu^2) g^{\mu\nu}+(3-2\, \mu^2) p_1^\mu p_2^\nu  \right\}_{\textrm{sym}} }_{\N^{(2)}_{\mu\nu}(\{p_i\},\mu^2)} q^\mu q^\nu \nonumber\\
  &+ \underbrace{\left\{4\, \mu^4 - 2\, p_1\cdot p_2\, \mu^2 +  m_X^4 \right\}}_{\N^{(0)}(\{p_i\},\mu^2)} 
\end{flalign}
where $\{p_i\}$ denotes the collective kinematic information of the phase-space point considered (external momenta, masses, helicity and color assignation, etc\ldots).
The $\{\}_{\textrm{sym}}$ notation indicates that the corresponding tensor components are symmetrized according to the procedure described by Eq.~\ref{eq:tensorpoly}.
The four-dimensional tensor numerator is thus obtained by setting $\mu^2=0$ and identifying the Lorentz structures multiplying the loop momentum
\begin{flalign}
 \N(q) = \N(q^2,\mu^2)\Big|_{\mu^2=0} ={}&  \underbrace{\left\{ 4g^{\mu\nu} g^{\rho\sigma}+2 p_1^\mu p_2^\nu g^{\rho\sigma}\right \}_{\textrm{sym}}  }_{\N^{(4)}_{\mu\nu\rho\sigma}(\{p_i\})} q^\mu q^\nu q^\rho q^\sigma \nonumber\\
  +&\underbrace{\left\{ 2 p_1\cdot p_2 g^{\mu\nu}+3 p_1^\mu p_2^\nu \right\}_{\textrm{sym}} }_{\N^{(2)}_{\mu\nu}(\{p_i\})} q^\mu q^\nu + \underbrace{\left\{m_X^4\right\}}_{\N^{(0)}(\{p_i\})}.
\end{flalign}
Several techniques have been proposed for constructing the tensor loop
numerator.  \MadLoop, which features an independent implementation of the {\sc\small OpenLoops} method~\cite{Cascioli:2011va}, progressively builds the loop numerator polynomial by successive calls to building block functions numerically computing the loop four-momentum dependence of each vertex and propagator involved in the loop. Contrary to analytic methods, this approach makes it very difficult to reconstruct the $d$-dimensional $\mu^2$ dependence of the loop numerator, so that it is important that the loop reduction algorithm works equally well with only the 4-dimensional projection $\N(q)$, in which case the missing rational terms arising from $\mu^2$ will be reconstructed independently.  An alternative method proposed in ref.~\cite{Heinrich:2010ax}, addressing the case where a numerical evaluation of the integrand is available but its full polynomial structure is not known, reconstructs the entries of the tensor by sampling the numerator on several values of the loop momentum.

In this paper, we will consider a one-loop tensor integrand to be defined by the entries of the
symmetric tensor numerator in Eq.~\eqref{eq:symtensornum}, as well as the momenta
$p_i$ and masses $m_i$ appearing in the loop denominators as in
Eq.~\eqref{eq:loopden}. Thanks to the new projection techniques introduced in this paper, these are now the only input required by \Ninja\ for performing the corresponding loop reduction.

\section{Semi-numerical integrand reduction via Laurent expansion}
\label{sec:semi-numer-integr}

In this section we briefly review the input needed by the semi-numerical
integrand reduction via Laurent expansion
algorithm~\cite{Mastrolia:2012bu} as implemented in the \textsc{C++}
library \textsc{Ninja}~\cite{Peraro:2014cba}.  We make no attempt in
giving a comprehensive description of this reduction method, which can
instead be found in ref.s~\cite{Mastrolia:2012bu,Peraro:2014cba}.
Indeed, in this paper we are merely interested in illustrating how to
provide the required input starting from a tensor integrand defined as
in sect.~\ref{sec:tensorintegrands}, while the internals of the
reduction algorithm are unchanged with respect to what has already
been presented in the literature.

Integrand reduction methods compute loop integrals by exploiting the
knowledge of the algebraic structure of the respective integrands.
In more details, any one-loop integrand in dimensional regularization
can be written as a sum of contributions with five or less loop
propagators, regardless of the number of external legs or the
complexity of the process.  The corresponding numerators, also known
as \emph{residues}, are polynomials with a universal,
process-independent parametric form.  The unknown process-dependent
coefficients appearing in this parametrization can thus be found via a
polynomial fit.  After integration, the amplitude is expressed as a
linear combination of known Master Integrals, whose coefficients can
be identified with a subset of the coefficients appearing in the
integrand decomposition.

An efficient way of computing the coefficients of the integrand
decomposition is by evaluating the integrand on \emph{multiple cuts},
i.e.\ on values of the loop momentum such that a subset of loop
denominators vanish.  Indeed, when evaluating the integrand on the
generic multiple cut $D_{i_1}=\cdots = D_{i_k}=0$, the only
non-vanishing contributions to the integrand decomposition are those
coming form the residues sitting on the cut (i.e.\ vanishing) denominators,
as well as from the higher-point residues having the cut denominators
as a subset of their propagators.  This suggested the possibility of
computing the coefficients of each residue by evaluating the integrand
on a subset of the solutions of the multiple-cut equations defined by
its loop denominators, subtracting the non-vanishing contributions form
higher-point residues and solving a system of equations for the unknown
coefficients.  This is therefore a top-down approach, where
higher-point residues are computed first (starting form 5-point
residues) and systematically subtracted from the integrand when
evaluating lower-point residues.  These are known as
\emph{subtractions at the integrand level}.  This is the approach
followed by the public reduction codes
\textsc{CutTools}~\cite{Ossola:2007ax} and
\textsc{Samurai}~\cite{Mastrolia:2010nb}.

As we mentioned, the integrand reduction via Laurent expansion
method~\cite{Mastrolia:2012bu} can achieve better stability and
performance by exploiting the knowledge of the analytic dependence of
the integrand on the loop momentum.  More specifically, on top of the
numerical evaluation of the loop numerator, the algorithm needs as
input a numerical evaluation of the leading terms of the numerator
with respect to properly defined Laurent expansions, parametric in the
directions of the loop momentum unconstrained by the multiple-cut conditions.  From
these, the coefficients of the integrand decomposition are computed
via a simplified polynomial division algorithm between the expansion
of the numerator and the loop denominators and corrected by
counter-terms depending on higher-point residues.  These are referred
to as \emph{subtractions at the coefficient level}, which simplify and
replace the ones at the integrand level of the original algorithm.

In the following we describe the four inputs needed by
\textsc{Ninja}, assuming the rank $R$ satisfies $R \leq n+1$, while in
the next sections we will describe how to automatically generate them
at run-time from the coefficients of a tensor numerator.  In this
section we consider a generic $\mu^2$-dependent numerator for the sake of
generality, although, as already mentioned, we will later specialize to the
case of a four-dimensional tensor numerator defined as in
Eq.~\eqref{eq:symtensornum}.  Notice however that the
$\mu^2$-dependence arising from the expansion of the loop momentum $q$
must be considered in both cases.  The four input functions are
\begin{itemize}
\item the \emph{numerator function} used for the cut-constructible part of 4-point residues and optional internal tests
  \begin{equation}
    \label{eq:numfun}
    \N(q,\mu^2),
  \end{equation}
  as a function of the loop momentum (notice that this is the same input as in
  traditional integrand reduction algorithms),
\item the $\mu^2$-expansion used for the rational part of 4-point residues, returning the terms $n_{\mu^2}^{(i)}$ defined by the expansion
  \begin{equation}
    \label{eq:muexp}
    \N(q,\mu^2) \Big|_{ q^\nu \rightarrow t\, v_0^\nu + v_1^\nu, \;  \mu^2 \rightarrow t^2\, v_0^2} \; \overset{t\rightarrow\infty}{=}\; n_{\mu^2}^{(R)}\, t^R+n_{\mu^2}^{(R-1)}\, t^{R-1} + \O(t^{R-2}),
  \end{equation}
 as a function of the four-vectors $v_0^\nu$ and $v_1^\nu$
 \begin{equation}
    \label{eq:muexpfun}
    n_{\mu^2}^{(i)} = n_{\mu^2}^{(i)}(v_j),
  \end{equation}
  
\item the $t_3$-expansion used for 3-point and 1-point residues, returning the terms $n_{t_3}^{(i,j)}$
  defined by the expansion
  \begin{equation}
    \label{eq:t3exp}
    \N(q,\mu^2) \Big|_{ q^\nu \rightarrow v_0^\nu + t\, v_3^\nu + \frac{\beta+\mu^2}{t} v_4^\nu} \; \overset{t\rightarrow\infty}{=}\; \sum_{i=R-4}^R \sum_{j=0}^{\lfloor (R-i)/2\rfloor} n_{t_3}^{(i,j)}\, t^i\, \mu^{2j} + \O(t^{R-5}),
  \end{equation}
 as a function of the four-vectors $v_0^\nu,v_3^\nu,v_4^\nu$ and the
 complex number $\beta$
 \begin{equation}
   \label{eq:t3expfun}
   n_{t_3}^{(i,j)} = n_{t_3}^{(i,j)}(v_0,v_3,v_4,\beta),
 \end{equation}
\item the $t_2$-expansion used for 2-point residues, returning the terms $n_{t_2}^{(i,j,k)}$
  defined by the expansion
  \begin{align}
    \label{eq:t2exp}
    & \N(q,\mu^2) \Big|_{ q^\nu \rightarrow v_1^\nu + x v_2^\nu + t\, v_3^\nu + \frac{\beta_0+\beta_1 x +\beta_2 x^2+\mu^2}{t} v_4^\nu} \; \overset{t\rightarrow\infty}{=}\; \nn
   & \qquad {}\sum_{i=R-3}^R \sum_{j=0}^{R-i} \sum_{k=0}^{\lfloor (R-i-j)/2 \rfloor} n_{t_2}^{(i,j,k)}\, t^i\, x^j \, \mu^{2k} + \O(t^{R-4}),
  \end{align}
 as a function of the four-vectors $v_1^\nu,v_2^\nu,v_3^\nu,v_4^\nu$ and the
 complex numbers $\beta_i$, with $i=0,1,2$
 \begin{equation}
   \label{eq:t2expfun}
   n_{t_2}^{(i,j,k)} = n_{t_2}^{(i,j,k)}(v_1,v_2,v_3,v_4,\beta_i).
 \end{equation}
\end{itemize}
We remind the reader that the vectors $v_{i}^\nu$ defining the expansions are cut-dependent, 
so that the methods for the corresponding coefficients will be called on all the relevant cuts (and possibly more than once per cut, as needed) within one loop reduction.
The terms above are all those needed for calculations with $R\leq n+1$.
If the rank is lower than $n+1$, fewer terms are needed, and in
numerical implementations one should take care that only a minimal
number of terms is computed so as to optimize performances.

Any one-loop generator capable of providing a numerical evaluation for
the terms in Eq.~\eqref{eq:muexpfun}, \eqref{eq:t3expfun}
and~\eqref{eq:t2expfun}, on top of the evaluation of the numerator as
in Eq.~\eqref{eq:numfun}, can use \textsc{Ninja}.  We now turn to describing a
method for building these expansions from a tensor numerator of the form
of~\eqref{eq:symtensornum}.  The algorithm then proceeds in a
purely numerical way, using as input only the (numerical) entries of a
symmetric tensor numerator.  Indeed, as already mentioned, the terms of
each expansion~\eqref{eq:muexp}-\eqref{eq:t2exp} can be defined as contractions between the tensor
numerator and cut-dependent tensors which \textsc{Ninja} can build at
runtime by means of recursive algorithms.  Since all the methods
defined above have been implemented within the \textsc{Ninja} library
for generic tensor integrands with $R\leq n+1$, this allows current one-loop
tensor generators to use \textsc{Ninja} for the reduction, simply by
providing the coefficients $\N^{(r)}_{\mu_1\cdots \mu_r}$ defining the
loop numerator as in Eq.~\eqref{eq:symtensornum}.

\section{Symmetric tensors}
\label{sec:symmetric-tensors}

In this section we introduce some notation on symmetric tensors and recursive formulas useful for efficiently building the
cut-dependent tensorial projectors appearing in our results.

\subsection{Notation}
\label{sec:notation}

Consider a set of independent vectors $\{ v_1\ldots v_k \}$ and the
symmetrized tensor product of $r$ (not necessarily distinct) vectors
$v_{i_1},\ldots , v_{i_r}$, with $i_1,\ldots,i_r\in \{1,\ldots,k\}$,
namely
\begin{equation*}
  \sum_{\sigma\in S_r} v_{i_1}^{\mu_{\sigma(i_1)}}\cdots v_{i_r}^{\mu_{\sigma(i_r)}}.
\end{equation*}
This tensor, being completely symmetrized, only depends on the number
of times each $v_i$ enters the product.  As noted in
ref.~\cite{comon:hal-00327599}, one can exploit this and introduce a
natural correspondence between symmetric tensors and polynomials.
More in detail, we will use the following polynomial notation
\begin{equation} \label{eq:tensorpoly}
  v_1^{r_1}\cdots v_k^{r_k} \equiv \frac{1}{r_1!\cdots r_k!}\, \sum_{\sigma\in S_r} v_{i_1}^{\mu_{\sigma(i_1)}}\cdots v_{i_r}^{\mu_{\sigma(i_r)}},
\end{equation}
where $r_i$ is the multiplicity of occurrence of $v_i$ on the r.h.s.\
of the equation, with $\sum_i r_i = r$.  The conventional prefactor we
introduced on the r.h.s.\ exactly cancels out against the equivalent
permutations of the tensor indexes, which turns out to be particularly
convenient for our application.  This is better clarified with a
couple of explicit examples
\begin{align}
  v_1^r = {} & v_1^{\mu_1}\cdots v_1^{\mu_r} \nn
  v_1^{r-1} v_2 = {} & v_2^{\mu_1} v_1^{\mu_2}\cdots v_1^{\mu_r}+v_1^{\mu_1} v_2^{\mu_2} v_1^{\mu_3}\cdots v_1^{\mu_r} + \cdots +  v_1^{\mu_1} \cdots v_1^{\mu_{r-1}} v_2^{\mu_r}.
\end{align}
The notation is also useful for writing tensor relations in a compact
way.

As a shorthand, if $T$ and $U$ are symmetric tensors of identical rank $r$, we also define the contraction
\begin{equation}
  \label{eq:TUcontract}
  T(U) \equiv \sum_{\mu_1\leq \cdots \leq \mu_r}\, T^{\mu_1\cdots \mu_r} U_{\mu_1\cdots \mu_r},
\end{equation}
where the sum over repeated indices is restricted so as to be consistent with
the definition in Eq.~\eqref{eq:symtensornum}.

\subsection{Recursive formulas}
\label{sec:recursive-formulas}

Tensors can be built recursively by multiplying lower rank tensors
with vectors.  For this purpose, we can define the \emph{tensor
  product} of a rank-$(r-1)$ tensor $T$ with a vector $v$ as the
rank-$r$ tensor
\begin{equation}
  \label{eq:tensorprod}
  T \otimes v \equiv T^{\mu_1\cdots \mu_{r-1}}\, v^{\mu_r}.
\end{equation}
Notice that, even if $T$ is symmetric, the r.h.s.\ will in general not
be a symmetric tensor.  However one can easily work out
\emph{recursive formulas} which build symmetric tensors from linear
combinations of tensor products.

The easiest recursive formula involves rank-$r$ tensors obtained by
multiplying a single vector $v$ with itself $r$ times, namely
\begin{equation}
  \label{eq:tenvr}
  v^r = v^{r-1} \otimes v.
\end{equation}

The easiest non-trivial case involving two vectors $v_1$ and $v_2$ is
\begin{equation}
  \label{eq:tenv1rv2}
  v_1^{r-1} v_2 = v_1^{r-1}\otimes v_2 + (v_1^{r-2} v_2)\otimes v_1.
\end{equation}
The tensor $v_1^{r-1}$ in the first addend on the r.h.s.\ can in
turn be built beforehand using Eq.~\eqref{eq:tenvr}, while the tensor
$v_1^{r-2} v_2$ appearing in the second addend is instead of the same
type of the one on the l.h.s.\ but with a lower rank.
Eq.~\eqref{eq:tenv1rv2} can thus be read as a recursive formula for
building symmetric tensors of the form $v_1^{r-1} v_2$, where the
recursion goes from lower to higher ranks $r$, starting from $r=1$
which trivially reads $v_1^0 v_2=v_2$.  A useful generalisation of
Eq.~\eqref{eq:tenv1rv2} is
\begin{equation}
  \label{eq:tenv1rmkv2k}
  v_1^{r-k} v_2^k = (v_1^{r-k}\otimes v_2^{k-1})\, v_2 + (v_1^{r-1-k} v_2^k)\otimes v_1,
\end{equation}
which should be read as a recursion relation in both $r$ and $k$.
Indeed, the second addend involves a symmetric tensor
$v_1^{r-1-k} v_2^k$ which is of the same type as the l.h.s.\ but with
total rank $r-1$.  The tensor in the first addend may instead be
rewritten as $v_1^{r-k}\otimes v_2^{k-1}=v_1^{r'-k'}\otimes v_2^{k'}$,
with $r'=r-1$ and $k'=k-1$, hence as a tensor of the same form as the
l.h.s.\ but with lower values of $r$ and $k$.

For our purposes we need one more recursive formula involving three
vectors $v_1$, $v_2$ and $v_3$, which reads
\begin{equation}
  \label{eq:tenv1v2v3}
  v_1^{r-k-1} v_2^k v_3 = (v_1^{r-k-1} v_2^k) \otimes v_3 + (v_1^{r-k-1} v_2^{k-1} v_3) \otimes v_2 + (v_1^{r-k-2} v_2^k v_3) \otimes v_1.
\end{equation}
The ingredients for this recursion are, similarly as before, tensors
with lower $r$ and $k$, as well as tensors of the form of
Eq.~\eqref{eq:tenv1rmkv2k}.

It is also worth observing that all these recursive formulas can be seen as special cases of a more general one
\begin{align}
  \label{eq:5}
  v_1^{k_1}\cdots v_n^{k_n} ={}& (v_1^{k_1-1} v_2^{k_2} \cdots v_n^{k_n}) \otimes v_1 \nn
&  + \cdots \nn
& +  (v_1^{k_1} \cdots v_{i-1}^{k_{i-1}}  v_i^{k_i-1} v_{i+1}^{k_{i+1}}\cdots v_n^{k_n}) \otimes v_i \nn
& + \cdots \nn
& + (v_1^{k_1} \cdots v_{n-1}^{k_{n-1}} v_n^{k_n-1})\otimes  v_n.
\end{align}

An important observation for numerical calculations is that these
recursive formulas have the nice side effect of automatically
embedding a system of abbreviations based on reusing common
subexpressions.  Indeed, as one can see from the definition in
Eq.~\eqref{eq:tensorprod}, each entry in a tensor product of total
rank $r$ can be obtained from an entry of rank $r-1$ by a single
multiplication.  Because our formulas are recursive on the rank and
involve linear combinations of tensor products, they provide a built-in
mechanism for reusing subexpressions of lower rank when building
tensors of higher rank.  Moreover, the possibility of reusing common
subexpressions is not limited to contributions defined within the same
recursive formula, but it can also be extended to contributions across
different equations in a way which fits particularly well with the
method we will use for building the Laurent expansions of the
integrands.  We will
see in the next section that the leading term of a Laurent expansion
can always be obtained from tensors of the form of
Eq.~\eqref{eq:tenvr}.  Next-to-leading terms, when needed, will be
constructed using Eq.~\eqref{eq:tenv1rv2}.  
As we already observed,
the r.h.s.\ of this equation involves a lower-rank tensor of the same
form as the l.h.s.\ and a tensor of the same form of
Eq.~\eqref{eq:tenvr}.  While the former is available simply by
implementing the recursion from lower to higher ranks, the latter can
instead be reused from the tensors recursively built for the leading
term.  An analogous strategy is also possible for Laurent expansion
terms beyond next-to-leading, where one can always use tensors built
in previous steps of the calculation as input for the recursive
formulas (more explicit examples will be given in
sect.~\ref{sec:tens-proj-laur}). This greatly reduces the total number of
operations needed for the construction of these tensors, without
hard-coding complex analytic formulas and while having a relatively
simple bookkeeping and still being completely general with respect to
the rank of the tensors appearing in the recursion relations.

In the next section we show how the Laurent-expansion terms needed by
\textsc{Ninja} can be generated by contracting tensor numerators with
tensors of the same kind as those in Eq.~\eqref{eq:tenvr},
\eqref{eq:tenv1rmkv2k} and~\eqref{eq:tenv1v2v3}.

\section{Tensor projectors for Laurent-expansion terms}
\label{sec:tens-proj-laur}
As we stated above, we can build the terms of the Laurent expansions
described in sect.~\ref{sec:semi-numer-integr} by contracting the
tensor numerator with appropriate cut-dependent tensors, which can
be seen as projectors.  These can in turn be built recursively
using the formulas of sect.~\ref{sec:recursive-formulas}.  We will
illustrate the method by explicitly working out a few cases.  A
complete list of formulas for all the tensor projectors is given in
Appendix~\ref{app:expl-results-expans}.

As stated above, in the following we will consider a four-dimensional tensor numerator defined as in Eq.~\eqref{eq:symtensornum}, which thus only depends on the four-dimensional components $q^\mu$ of the loop momentum.  However, because \textsc{Ninja} implements a $d$-dimensional version of the integrand reduction method, the parametrization of $q$ on the $d$-dimensional cut solutions (and thus its Laurent expansion) will still depend on the extra-dimensional variable $\mu^2$ and we must therefore keep track of this dependence while building the expansions (knowing, of course, that it can only come from the loop momentum, and not from the numerator).  This $d$-dimensional reduction yields, on top of the coefficients of the master integrals, the contributions to the rational part of the amplitude coming from the $\mu^2$ dependence of the loop denominators, also known as $R_1$.  An alternative approach to the calculation of $R_1$ (used for instance by \textsc{CutTools}) is its reconstruction from the coefficients of a purely four-dimensional reduction, also known as \emph{cut-constructible} part.  It is worth stressing that, as one can observe from the results collected in Appendix~\ref{app:expl-results-expans}, the calculation of the Laurent expansion terms involving $\mu^2$, with the approach presented in this paper, can always be recycled from the identical terms needed for the cut-constructible part, except for the box residues where the expression is however very simple.
This allows to efficiently provide the algorithm of \textsc{Ninja} with all the terms needed by its $d$-dimensional integrand reduction, while remaining completely agnostic about the $\mu^2$-dependence of the loop numerator within the reduction routines. 

In the following, $n$ is the number of loop propagators and $R$ is the
rank of the tensor numerator.  The Laurent-expansion parameter is
denoted by $t$, and it is always convenient to compute the terms from
higher to lower powers of $t$.  A first reason for this is that the
highest powers of $t$ are always needed, while the lower powers might
not be.  A second compelling reason is that, as we already mentioned,
tensors built for the leading terms in $t$ can be reused as input for
building terms with lower powers of $t$, using the recursive formulas
introduced in the previous section.

\subsection{Numerator evaluation}
\label{sec:numerator-evaluation}

The easiest function to provide is the evaluation of the numerator function which, using Eq.~\eqref{eq:tensornum}, simply amounts to
\begin{equation}
  \label{eq:6}
  \N(q) = \sum_{r=0}^R \N^{(r)}(q^r),
\end{equation}
where we used the notation introduced in
sect.~\ref{sec:symmetric-tensors} (in particular
Eq.~\eqref{eq:tensorpoly} and~\eqref{eq:TUcontract}).  Each tensor $q^r$ can in turn be
built using the recursive formula of Eq.~\eqref{eq:tenvr}, which here
reads
\begin{equation}
  q^r = q^{r-1} \otimes q.
\end{equation}

\subsection{The $\mu^2$-expansion}
\label{sec:mu2-expansion}
The $\mu^2$-expansion, only needed for $R\geq n$, involves a single
term for $R=n$ and two terms for $R=n+1$.  Since the vector $v_0$
comes with a power of $t$, while $v_1$ is $\O(t^0)$, it is
straightforward to see that the leading term in $t$ of the expansion
defined by Eq.~\eqref{eq:muexp} is
\begin{equation}
  \label{eq:8}
  n_{\mu^2}^{(R)} = \N^{(R)}(v_0^R).
\end{equation}
As done before, we can build $v_0^R$ recursively by means of Eq.~\eqref{eq:tenvr}
\begin{equation}
  \label{eq:tenv0r}
  v_0^r = v_0^{r-1} \otimes v_0.
\end{equation}
For the case where $R=n+1$, we also need the next-to-leading term in
$t$, which is given by the following formula
\begin{equation}
  \label{eq:9}
  n_{\mu^2}^{(R-1)} =  \N^{(R-1)}(v_0^{R-1}) + \N^{(R)}(v_0^{R-1} v_1),
\end{equation}
where the tensor appearing in the first addend of the r.h.s.\ was
already built for the leading terms using Eq.~\eqref{eq:tenv0r}, while
for the second addend we can use the recursive relation in Eq.~\eqref
{eq:tenv1rv2}, which in this case reads
\begin{equation}
  v_0^{r-1} v_1 = v_0^{r-1}\otimes v_1 + (v_0^{r-2} v_1)\otimes v_0.
\end{equation}
The first addend of this recursive relation also depends on tensors
built using Eq.~\eqref{eq:tenv0r} for the leading term, while the
second depends on a lower-rank tensor which gives the recursion in
$r$.

\subsection{The $t_3$-expansion}
\label{sec:t_3-expansion}
The $t_3$-expansion has a more complicated structure due to the
presence of three vectors ($v_0$, $v_3$ and $v_4$) and the free
variable $\mu^2$ on top of the expansion variable $t$.  More in detail,
the vector $v_3$ comes with a power of $t$, the vector $v_0$ is
$\O(t^0)$ and the vector $v_3$ has a $\O(1/t)$ term multiplied by
the constant $\beta$ and a $\O(\mu^2/t)$ term.  Hence, the projector
for the leading term is a tensor containing only $v_3$, while
replacing a $v_3$ by a $v_0$ decreases the power in $t$ by one, and
replacing a $v_3$ by a $v_4$ decreases the power in $t$ by two and
also adds a $\mu^2$ term.

Since the leading and next-to-leading terms of the expansion do not
involve $v_4$ (and thus neither $\mu^2$), they have exactly the same structure
as those for the $\mu^2$-expansion.  They are
\begin{equation}
  n_{t_3}^{(R,0)} = \N^{(R)}(v_3^R).
\end{equation}
and
\begin{equation}
  \label{eq:9}
  n_{t_3}^{(R-1,0)} =  \N^{(R-1)}(v_3^{R-1}) + \N^{(R)}(v_3^{R-1} v_0).
\end{equation}

The next-to-next-to-leading terms in $t$ are two, namely a
$\O(t^{R-2}\mu^0)$ term and a $\O(t^{R-2}\mu^2)$ term, and their
expression involves all the three vectors $v_0$, $v_3$ and $v_4$.
They are given by the following formulas
\begin{align}
  n_{t_3}^{(R-2,0)} ={}& \N^{(R-2)}(v_3^{R-2}) + \N^{(R-1)}(v_3^{R-2} v_0)  + \N^{(R)}(v_3^{R-2} v_0^2) + \beta\, \N^{(R)}(v_3^{R-1} v_4) \\
  n_{t_3}^{(R-2,1)} ={}& \N^{(R)}(v_3^{R-1} v_4).
\end{align}
It is worth making a few observations.  We already mentioned that the
$\mu^2$-dependent terms can be determined from the cut-constructible
ones, and indeed the contribution $\N^{(R)}(v_3^{R-1} v_4)$ is common
between the two equations and thus only needs to be computed once.
Moreover the tensor $v_3^{R-1} v_4$, and more in general all those of
the form $v_3^{r-1} v_4$, can be computed from the recursion relation
in of Eq.~\eqref{eq:tenvr} which in this case depends on the tensors
$v_3^r$ already computed above for the leading term in $t$.  The
tensor $v_3^{R-2} v_0^2$ can instead be computed using the formula in
Eq.~\eqref{eq:tenv1rmkv2k} with $k=2$, to be read as a recursion
relation in $r$ and depending on tensors of the form $v_3^{r-1} v_0$
already computed for the next-to-leading terms in $t$.

Similarly, one can compute all the other terms with lower powers of
$t$, by means of a simple power counting on the vectors $v_0$, $v_3$
and $v_4$, and building appropriate tensors with the formulas of
sect.~\ref{sec:recursive-formulas}.  One can also check that the
formulas of Eq.~\eqref{eq:tenvr}, \eqref{eq:tenv1rmkv2k}
and~\eqref{eq:tenv1v2v3} suffice for the calculation of all the terms
down to $\O(t^{R-4})$, which is all one needs for integrands with
$R\leq n+1$.  At each step one can perform the recursion with respect
to $r$, for $r=0,\ldots,R$, reusing as ingredients tensors computed
for lower $r$ or for terms with higher powers of $t$.  Explicit
formulas for all terms are given in
Appendix~\ref{app:expl-results-expans}.

\subsection{The $t_2$-expansion}
\label{sec:t_2-expansion}

The terms for the $t_2$-expansion can be computed using the same
method as for the $t_3$-expansion.  Since we are now dealing with
three variables ($t$, $x$ and $\mu^2$) and four independent vectors
($v_i$, with $i=1,\ldots,4$) the main difference is a more involved
bookkeeping, partially mitigated by the need of less powers in $t$.
Explicit formulas are collected in
Appendix~\ref{app:expl-results-expans}.

\section{Implementation}
\label{sec:implementation}

The tensor projectors for the Laurent expansion terms described above
have been implemented in the \textsc{Ninja} library.

The reduction algorithm implemented in \textsc{Ninja} requires as
input a numerator which is an abstract interface implementing the
methods described in sect.~\ref{sec:semi-numer-integr} (the
\textsc{C++} programming interface is described in
ref.~\cite{Peraro:2014cba}).  We thus implemented such an interface
which computes the expansion terms collected in
Appendix~\ref{app:expl-results-expans} from the coefficients
$\N^{(r)}_{\mu_1\cdots \mu_r}$ of a generic tensor numerator, defined
according to Eq.~\eqref{eq:symtensornum}.

The tensor numerator is treated as a polynomial whose coefficients are stored in a unidimensional array, from lowest to highest according to the \emph{graded lexicographic monomial order} in the variables $q^\mu$ with $q^0\prec q^1 \prec q^2 \prec q^3$ (i.e.\ terms are ordered by their total degree and terms with the same total degree are ordered lexicographically).  This is the same monomial order used internally by \textsc{MadLoop} and turns out to be particularly convenient for building the tensors described in this paper, since we use formulas which are recursive with respect to the total rank.  It is worth observing however that none of the results presented in this paper rely on a specific representation of the momenta (and consequently of the tensor numerator).  In particular, the formulas collected in Appendix~\ref{app:expl-results-expans}, as well as the algorithms implemented for building the corresponding tensors, are unchanged after a change of coordinates $q^\mu\to q'^\mu=\Lambda^\mu{}_\nu\ q^\nu$ and can thus be applied to any other representation of the four-dimensional components after converting the momenta $v_i$ used as input into the alternative representation.

We also implemented in \textsc{Ninja} a \textsc{Fortran-90} wrapper exposing this tensor interface, which in principle can be used by any one-loop tensor-based generator by specifying the loop propagators and the coefficients of the tensor numerator defining the integral to be computed.  Both the \textsc{Fortran} and the \textsc{C++} interface are publicly available since version 1.1.0 of the library.

\MGaMC\ (v2.3.4 and onwards) now includes a version of \Ninja\ which is used as the default loop reduction method\footnote{Other alternative reduction algorithms (\CutTools, \PJFry, \Golem, \IREGI\ and \Samurai) are still used in the event that \MadLoop\ stability tests deem the kinematic configuration numerically unstable.}. This default installation of \textsc{Ninja} can be automatically updated to the latest online one (independently distributed) by running the following command in the \MGaMC\ interactive interface:

\noindent\\
~\prompt\ {\tt ~install ninja}\\

\Ninja, similarly to \CutTools\, is available both in double and quadruple precision at runtime and \MadLoop\ will dynamically switch to quadruple precision when its internal stability tests indicate that the double precision computation does not meet the requirement in numerical accuracy.
In \MadLoop, the computation of the tensor numerator coefficients is completely independent from the loop reduction and, as a result, the stability rescue is first attempted by re-performing in quadruple precision \emph{only the loop reduction} (although with input kinematics already promoted to quadruple precision accuracy). This is often enough to restore satisfactory numerical stability, hence avoiding the much more time-consuming full-fledged quadruple precision computation.

We also point out that \Ninja's internal kinematic matrix
$K_{ij}={(p_i-p_j)^2}$, with quantities defined as in
Eq.~\eqref{eq:loopden}, is initialized directly in \MadLoop\ where the
following three on-shell limits are set to be exact when below a
certain adimensional threshold $\delta$ set to $10^{-8}$ by default:
\begin{itemize}
\item If $|m_{i(j)}|^2>0$, set $K_{ij} = 0$ if $\left| \frac{(p_i-p_j)^2-m_{i(j)}^2}{m_{i(j)}^2} \right| < \delta$
\item If $|E_i+E_j|>0$, set $K_{ij} = 0$ if $\left| \frac{2(p_i-p_j)}{E_i+E_j} \right| < \delta^2$
\end{itemize}
This proved to help the numerical stability of the reduction, essentially because it avoids ever approaching the kinematic region where the master integrals switch from a massive to massless description.
The choice of the analytic expression to be evaluated by the Master Integral library is typically controlled by an internal infra-red threshold parameter which would apply to each integral independently. By regulating the kinematic matrix in \MadLoop, we guarantee the consistency of the expression employed for all master integrals.

Finally, all loop reduction methods except \CutTools\ and \IREGI\ can be independently (de-)activated before the generation of the one-loop matrix element numerical code by setting the corresponding \MGaMC\ path options in {\tt <MG\_root>/input/mg5\_configuration.txt}. If activated at generation time, then their use at run time can be controlled via the parameter {\tt MLReductionLib} specified in the file {\tt MadLoopParams.dat}.

\section{Applications}
\label{sec:applications}
In this section, we will present the summary of a detailed study of the timing and stability performances of \MadLoop\ interfaced to \Ninja. When available, we compare the results obtained with \Ninja\ against other reduction algorithms, namely \CutTools, \Samurai, \IREGI, \PJFry and \Golem, whose limitations (for the versions interfaced to \MadLoop) are summarized in table~\ref{tab:MLTIR}.  \CutTools\ is a library implementing a four-dimensional version of the integrand reduction method, as well as a reconstruction of the $R_1$ term as explained at the beginning of section~\ref{sec:tens-proj-laur}.  \Samurai\ is a similar tool which always performs a full $d$-dimensional integrand reduction, making it capable of handling $d$-dimensional loop numerators at the price of being less efficient of four-dimensional ones, since it implements a more complex reconstruction of the integrand.  \IREGI, \PJFry and \Golem\ are instead tensor integral reduction tools.
\begin{table}[h!]
\begin{center}
\begin{tabular}{C{3cm}C{5cm}C{5cm}}
 Reduction tool & Max. $n_\textrm{loop\_prop.}$ & Max. rank \\
 \midrule \
{\sc\small Ninja}  & unlimited & $n_\text{loop\_prop.}+1$ \\
{\sc\small Samurai}  & unlimited$^\star$ & $n_\text{loop\_prop.}+1$ \\
{\sc\small CutTools}   & unlimited$^\star$ & $(n_\text{loop\_prop.}+1)^\dagger$ \\
{\sc\small IREGI}       & 7 & $6^{\star\star}$ \\
{\sc\small PJFry}       & $5$ & $n_\text{loop\_prop}$ \\
{\sc\small Golem95}  & $6$ & $\max(6, n_\text{loop\_prop.}+1)$ \\
\multicolumn{3}{l}{}\\
\multicolumn{3}{l}{\small{$^\star$: For reducing loops with 9 (11) loop lines and more, \Samurai\ (\CutTools) must be recompiled }}\\
\multicolumn{3}{l}{\small{\phantom{$^\star$:l} with an increased value for its default maximum number of denominators.}}\\
\multicolumn{3}{l}{\small{$^\dagger$: Loops with rank $n_\text{loop\_prop.}+1$ are supported in {\sc\small CutTools} only for models with effective }}\\
\multicolumn{3}{l}{\small{\phantom{$^\dagger$:l} interactions involving only the Lorentz structures of the Higgs-gluons vertices.}}\\
\multicolumn{3}{l}{\small{$^{\star\star}$: This \IREGI\ limitation stems from the observation that its reduction of loops with rank }}\\
\multicolumn{3}{l}{\small{\phantom{$^{\star\star}$:l} larger than 6 is typically unstable for all kinematic configurations.}}
\end{tabular}
\end{center}
\caption{\label{tab:MLTIR} Limitations of the different reduction methods interfaced to \MadLoop. The notation $n_\text{loop\_prop.}$ refers to the number of internal propagators in the loop considered. All reduction tools except {\sc\small PJFry} support complex masses.}
\end{table}
We stress that \MadLoop\ can dynamically change at run time the active reduction tool depending on its applicability to each individual (group of) loop(s) being reduced. 

The study carried in this section focuses on the following five classes of processes, chosen for their different characteristics that cover a wide spectrum of one-loop matrix-element computations.
The notation $\{i,j\}\cdot X$ denotes that we considered all the processes with either $i$ or $j$ occurrences of particle $X$ in the final states.
\begin{itemize}
\item A) $g g \rightarrow t \bar{t} + \{ 0,1,2,3\}\cdot g$\\
This class of processes is a common benchmark for pure QCD computations as it introduces the top mass as an additional scale. The one-loop amplitudes for each multiplicity of this class of processes were first computed in ref.s~\cite{Nason:1987xz,Dittmaier:2007wz,Bevilacqua:2010ve}. The one-loop matrix element for the process $g g \rightarrow t \bar{t} g g g$ is generated and computed here for the first time for specific kinematic configurations (see Appendix.~\ref{sec:ggttxggg}).
\item B) $g g \rightarrow H + \{ 1,2,3\}\cdot g$\\
These processes are computed within the Higgs Effective Interaction
Theory as implemented in~\cite{Artoisenet:2013puc}.
In this effective theory the top-quark loop is integrated out, yielding effective interactions between gluons and the Higgs. The resulting dimension-5 operators lead to
loops with rank $n_\text{loop\_prop.}+1$ which are especially
challenging to reduce. Thanks to the trivial Lorentz structure of the effective Higgs interactions, both \CutTools\ and older versions of \Samurai\ are applicable~\cite{Mastrolia:2012bu}, even though they do no support completely general tensor numerators of higher rank. The one-loop amplitudes for each multiplicity of this class of processes were first computed in ref.s~\cite{deFlorian:1999zd,Cullen:2013saa,vanDeurzen:2013rv}.
\item C) $g g \rightarrow Y + \{ 1,2,3\}\cdot g$\\
This set of processes is similar to the one above, but involving a spin-2 particle $Y$ of mass 1 TeV and whose graviton-like effective interactions are described in sect.~2.3 of ref.~\cite{Artoisenet:2013puc} (we considered $\kappa_g=\kappa_q$). In this case, the tensor numerator of the resulting loops with rank $n_\text{loop\_prop.}+1$ can have an arbitrary structure that \CutTools\ cannot handle. The one-loop amplitudes for this class of processes were first computed in~\cite{Mathews:2004xp} and~\cite{Karg:2009xk} for 0 and 1 additional gluon in the final states and are computed here for the first time for 2 and 3 additional gluons (see Appendix.~\ref{sec:ggY23g}).
The study of the phenomenology of QCD corrections within this effective theory featuring a spin-2 particle is in preparation~\cite{GravitonInPrepatation}.
\item D) $g g \rightarrow \{2,3,4,5\}\cdot Z$\\
This is a class of loop-induced processes for which event generation has recently been automated in \MGaMC~\cite{Hirschi:2015iia}. Loop-induced processes are processes without any contribution from tree-level Feynman diagrams, in which case the one-loop amplitude must be squared against itself. This implies that when using integrand reduction (and only then), loops must be reduced individually and independently for each helicity configuration.

Also, given the absence of any Born contribution, loop-induced processes are finite and build by themselves the complete Leading-Order (LO) prediction. For this reason, the speed of event generation and phase-space integration is entirely driven by the one of the one-loop matrix element, making optimizations especially relevant in this case. The gluon-fusion amplitude for $g g \rightarrow Z Z$ was first computed in~\cite{Glover:1988rg} and results for the processes $g g \rightarrow Z \gamma \gamma$ and $g g \rightarrow Z Z Z$ were shown in~\cite{Agrawal:2012as,Hirschi:2015iia}, while the loop-induced processes with four and five final state Z-bosons have never been studied.

\item E) $u \bar{u} \rightarrow Z Z Z Z Z$, $u \bar{u} \rightarrow e^+ \nu_e \mu^-\bar{\nu}_{\mu} b \bar{b}$,
                                                        $u \bar{u} \rightarrow t \bar{t} b \bar{b} d \bar{d}$ and $u \bar{u} \rightarrow t \bar{t} b \bar{b} d \bar{d} g$ \\
This less uniform class of process with $u \bar{u}$ initial states
serves different purposes. The first process is intended to be compared
with its loop-induced counterpart. The second one includes both EW and
QCD loop contributions, of all coupling orders, and it probes the
behavior of the loop reduction algorithms in the presence of many
scales and with complex masses in the loop propagators. The last two
processes test the reduction for high multiplicity processes featuring
loops with a large number of loop propagators (up to nine\footnote{Note that contrary to \Ninja, for enneagons and above \Samurai\ must be recompiled after its hard-coded limit on the number of propagators has been increased.}) but low rank.
These four high multiplicity processes have been selected for their specific characteristics from the standpoint of loop reduction and they have no direct phenomenological relevance except for the second one, so that their computation is not present in the literature.

\end{itemize}

The b-quarks are considered massive in all SM processes except for $u \bar{u} \rightarrow e^+ \nu_e \mu^-\bar{\nu}_{\mu} b \bar{b}$.

\subsection{Timing profile}

For all processes listed at the beginning of this section, we measure
independently the time spent for the computation of the tensor components\footnote{This timing profile was performed \emph{prior to} an optimization in the computation of the tensor numerator (filtering out some tensor components which are analytically zero) which brings a minor improvement on \tnum\ of order $\mathcal{O}$(30\%) (process dependent). Note that, this does not affect the key comparison of the reduction time \tred\ between different reduction algorithms.} of the loop numerator (\tnum) and the reduction of the loops (\tred), for one random kinematic and helicity configuration summed over color assignations. We stress that \tred\ includes the time spent in evaluating the master integrals as well as in the computation of the coefficients of the Laurent expansion in the case of \Ninja.

In \MadLoop, there is no optimization in the computation of the loop integrand tensor numerator across helicity configurations, so that \tnum\ scales linearly with the number of non-vanishing helicity configurations. Conversely, \tred\ remains independent of that number since the summation over helicity configurations can be performed before the loop reduction (except for loop-induced processes when using integrand reduction techniques).

We stress that the timing profile for a single helicity configuration is the relevant figure of merit for applications within \MGaMC\ which does not explicitly sum over helicity configurations for loop contributions, but instead adopts a Monte-Carlo procedure coupled with an adaptive importance sampling.

\begin{figure}[h!]
	\centering
	\includegraphics[trim=80 108 45 70,clip,scale=0.7]{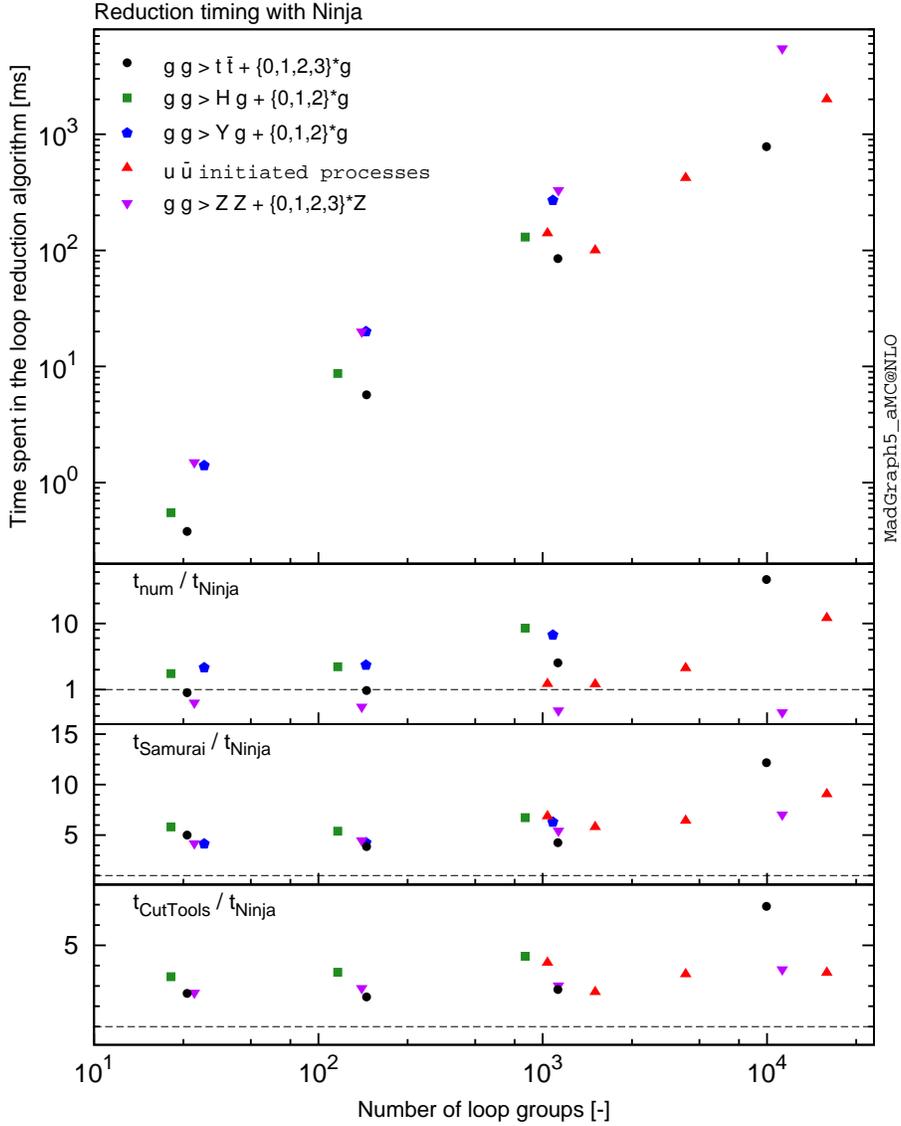}
	\caption{\label{fig:TimingPlot} Overview of the timing performances of \Ninja, \Samurai\ and \CutTools\ interfaced to \MadLoop\ on a single core of MacBook (OS 10.8.5), 2.7 GHz Intel Core i7 and using the GNU {\tt gfortran -O2} (v4.8.2) compiler. The timings refer to the computation of the one-loop matrix element summed over color assignations but for a \emph{single helicity} configuration. A loop group combines all loops which can be reduced together. Details on the processes considered are given at the beginning of sect.~\ref{sec:applications}.}
\end{figure}

We summarize our findings in fig.~\ref{fig:TimingPlot} showing results obtained with \MadLoop\ interfaced to either \Ninja, \Samurai\ or \CutTools. 

The x-axis registers the number of loop groups which combines all loops that can be reduced together. This corresponds to the set of loops sharing the same topology (i.e.\ ordered list of loop propagators), except for loop-induced processes where each loop must be reduced individually and therefore lies in a loop group of its own. Notice that since loops identical up to couplings (like fermion loops of different flavors) are combined already when generating the loop matrix element code, they only count as one.

The main feature of the upper panel of fig.~\ref{fig:TimingPlot} is that within each class of processes, the dependence of the reduction time w.r.t. the number of loop groups is linear, as already noticed in~\cite{Cascioli:2011va}. The offset between each class of process is related to the difference in the rank of the constituting loops. The loop rank is typically larger in processes within models with higher dimensional operators (blue hexagons and green squares) as well as in loop-induced processes which involve fermionic loops only (purple triangles). Conversely, the rank becomes smaller as the number of external fermion lines increases and we indeed observe that the timings for the processes $g g \rightarrow t \bar{t} + n\cdot g$ and $u \bar{u} \rightarrow t \bar{t} b \bar{b} d \bar{d} (g)$ sit on a line underneath (black circles and red triangles).
It is interesting to note that that the process $u \bar{u} \rightarrow 5 \cdot Z$ is almost two orders of magnitude faster than its loop-induced counterpart, even though they are both contributions to the same final state.

The second inset of fig.~\ref{fig:TimingPlot} shows the ratio of the time spent in the computation of the components of the tensor numerator with the loop reduction time with \Ninja.
This ratio rapidly increases with the multiplicity and number of loop groups, clearly establishing that within the \MadLoop+\Ninja\ implementation, the computation time is asymptotically dominated by the computation of the loop integrand numerator. When no loop grouping is possible, as it is the case for loop-induced processes, we observe the opposite asymptotic behavior hence showing the the loop grouping plays an essential role in this limit.
We remind the reader that these conclusions apply to the computation of the loop matrix element for a single helicity configuration and only in the context of \MadLoop's technique for the computation of the loop integrand, which is most flexible but less optimal than full-fledged recursion relations~\cite{Actis:2012qn}, especially for processes with large multiplicity.

The bottom two insets of fig.~\ref{fig:TimingPlot} compare the performances of the three integrand reduction techniques interfaced to \MadLoop\ and reveals that \Ninja\ is about 3 to 5 times faster than \CutTools\ and 5 to 10 times faster than \Samurai. The reduction time relative to \Ninja\ increases with the process multiplicity, hence assessing the impact of the advantages of the integrand reduction via the Laurent expansion method as the complexity of the considered processes increases.

We include in appendix~\ref{app:timing_table} a table detailing the timing profile presented in fig.~\ref{fig:TimingPlot} as well as process generation times and results obtained with the tensor integral reduction tools \IREGI, \PJFry\ and \Golem.

\subsection{Stability study}

We now turn to the assessment of the numerical stability of \Ninja\ for the benchmark processes A)-D) listed at the beginning of this section. We do so by applying the internal stability tests of \MadLoop\ to a set of $N_{PS}$ random kinematic configurations and we report the resulting accuracy as a cumulative distribution for the fraction of points with a reduction relative accuracy larger than some target $\Delta$ on the x-axis. 

For the $2\rightarrow n$ processes, we chose $N_{PS}$ to be 100K for $n$=2, 10K for $n$=3 and 1K for $n$ $>$ 3. 
These $N_{PS}$ kinematic configurations are chosen randomly, with the constraint that all final states satisfy $p_{t,i} > 50$ GeV with angular separation $\Delta R_{ij}=\sqrt{\Delta\phi_{ij}^2+\Delta\eta_{ij}^2}\;>\;0.5$.
The center of mass energy chosen is $1$ TeV, except for the processes involving the spin-2 particle $Y$  in which case the center-of-mass energy is set to $1.2$ TeV.

\MadLoop\ combines two stability tests to estimate the numerical accuracy of the result:
\begin{itemize}
\item \emph{Loop direction test}: The loop reduction is performed a second time with the order of all propagators reversed (corresponding to the loop momentum redefinition $q\rightarrow -q$) and compared to the original evaluation. This changes the internal numerics of the reduction, hence assessing its stability. Given that the input kinematics remains unchanged, the tensor numerator components do not have to be recomputed.
\item \emph{Lorentz test}: The input kinematic is transformed by a Lorentz rotation for which the loop-matrix element is recomputed and compared to the original one.
\end{itemize}
Another commonly-used kind of stability test consists in rescaling all dimensionful quantities by a common factor. This is not used by \MadLoop\ because it proves to be impractical in the general case where the dimension of each of the model input parameters is not necessarily available within the generated code.

The stability tests are performed on a computation of the loop matrix element summed over all helicity configurations, except for the loop-induced processes for which only the \emph{all-minus} helicity configuration is considered.

\begin{figure}[h!]
	\centering
	\includegraphics[trim=80 368 45 70,clip,scale=0.8]{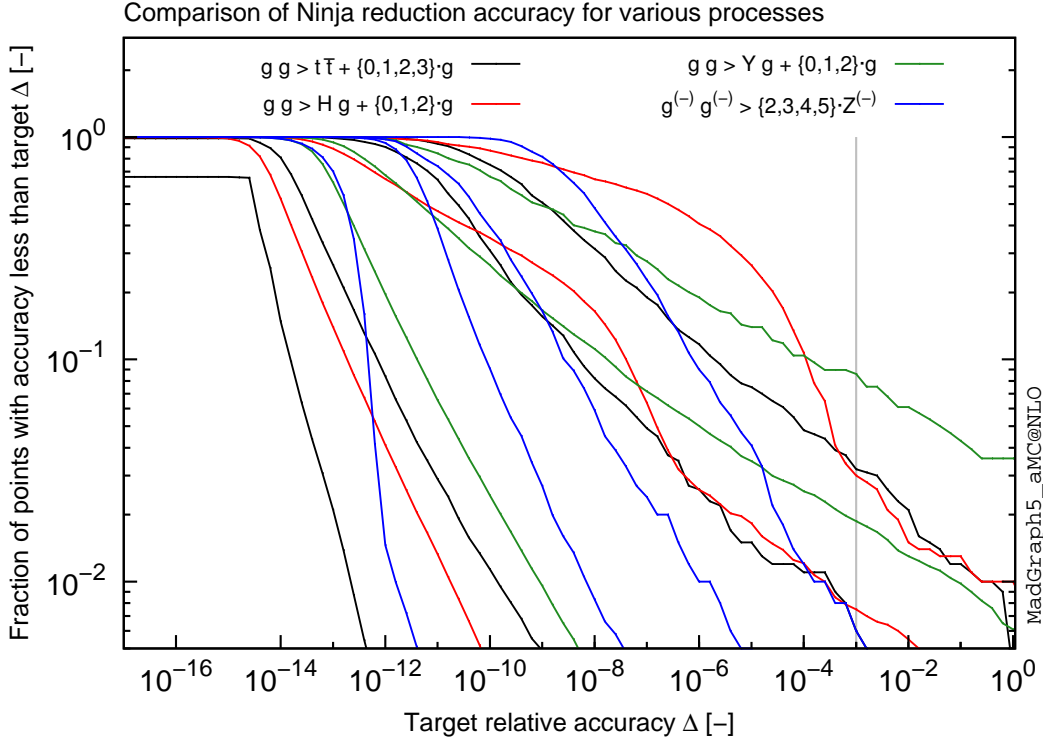}
	\caption{\label{fig:StabilityPlot} Comparison of the fraction of points with accuracy smaller than the target accuracy $\Delta$ on the x-axis obtained with \MadLoop+\Ninja\ (using double precision arithmetics) for a variety of processes (see text for details). The cumulative distributions shown are obtained from kinematic configurations with $\sqrt{s}=1$ TeV ($1.2$ TeV for the processes involving the spin-2 particle Y of mass $1$ TeV), randomly chosen with the constraints that all final states have a $p_{t,i} > 50$ GeV (except for the loop-induced processes) and an angular separation $\Delta R = \sqrt{\Delta \phi^2 + \Delta \eta^2} > 0.5$. The number of points considered is 100K, 10K, 1K and 1K for processes with 2, 3, 4  and 5 final states respectively. A vertical gray bar is shown at $\Delta=10^{-3}$ which corresponds to the typical threshold applied during event generation.}
\end{figure}

The vertical gray line at $\Delta=10^{-3}$ (i.e.\ 3 stable digits) marks the typical threshold used for Monte-Carlo event generation during which \MadLoop\ will attempt to rescue the phase-space points with a numerical stability estimate larger than this target by repeating the reduction (and possibly the computation of the tensor numerator) in quadruple precision.
The crossing of the various curves with this gray line therefore gives the fraction of unstable kinematic configurations for which this rescuing procedure will be necessary.
For all the processes of highest multiplicity in each class A)-D), this fraction is larger than 1\% and almost 10\% for $g g \rightarrow Y g g g$, which shows that numerical stability becomes an important issue\footnote{Faster and more flexible rescue mechanisms may be obtained using expansions around unstable kinematics, as implemented in the private tool \COLLIER~\cite{Denner:2005nn,Denner:2010tr}.} when attempting the integration of processes with loops of rank 6 and especially 7.

\begin{figure}[h!]
	\centering
	\includegraphics[trim=80 280 45 70,clip,scale=0.43]{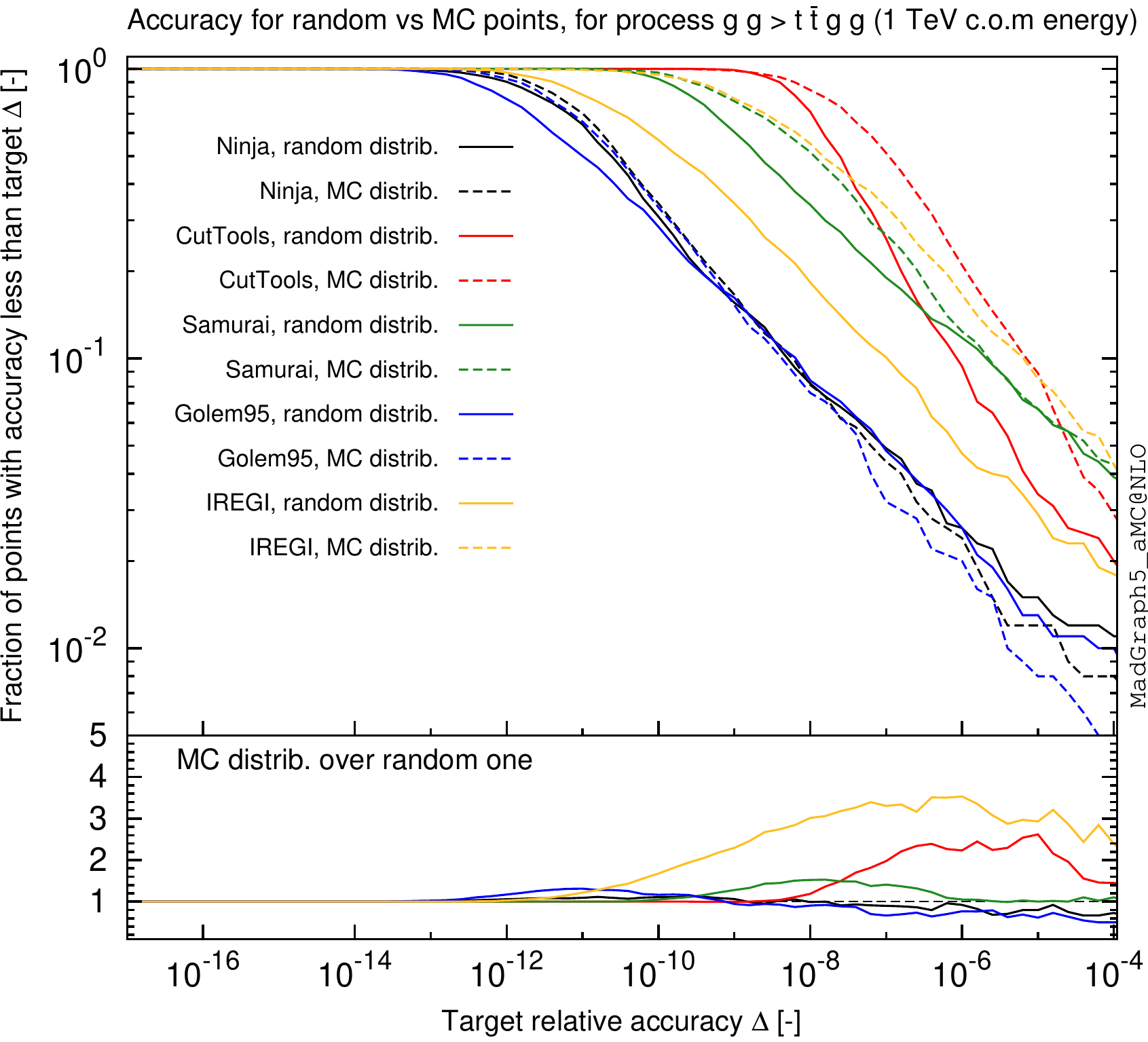}
	\includegraphics[trim=80 280 45 70,clip,scale=0.43]{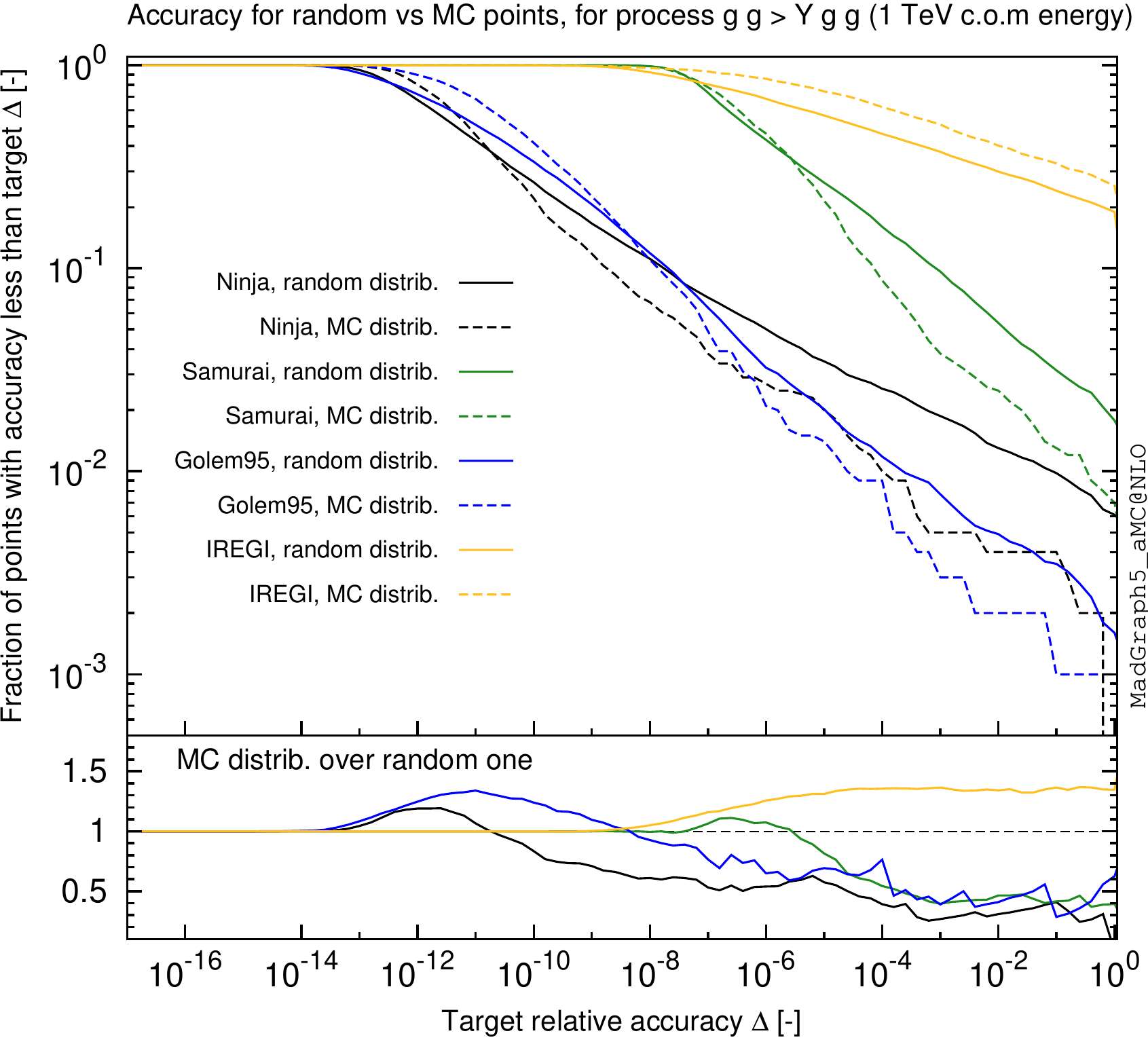}
	\includegraphics[trim=80 280 45 70,clip,scale=0.43]{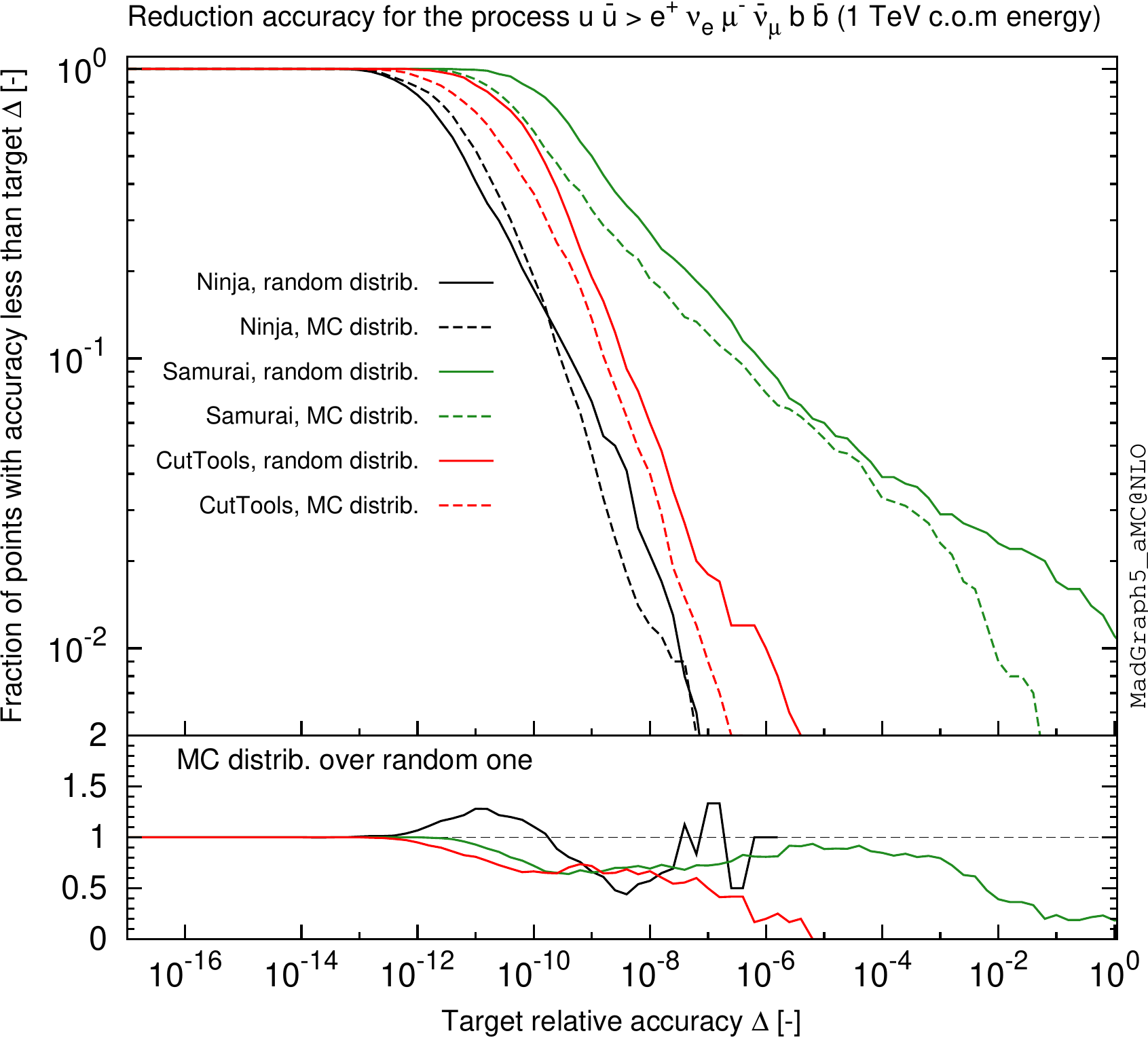}
	\caption{\label{fig:Stability_MC_vs_Random} The setup is identical to the one described in the caption of fig.~\ref{fig:StabilityPlot}. The stability profiles obtained from a random distribution of kinematic configurations are compared to the ones obtained from unweighted events generated with LO accuracy at LHC14, using the NNPDF 2.3 (NLO) PDF set.}
\end{figure}

Fig.~\ref{fig:Stability_MC_vs_Random} compares the stability of all applicable reduction tools for the processes $g g \rightarrow t \bar{t} g g$, $g g \rightarrow Y g g$ and $u \bar{u} \rightarrow e^+ \nu_e \mu^- \bar{\nu}_{\mu} b \bar{b}$. 
We observe that \Ninja\ is always the most stable reduction, comparable to that of \Golem\ which is however considerably slower (see appendix~\ref{app:timing_table}).

In fig.~\ref{fig:StabilityPlot}, comparing the relative position of the curves for processes with equal number of external legs shows that the determining factor for stability is the tensor numerator rank. This is also manifest when observing that despite the large multiplicity of the process $u \bar{u} \rightarrow e^+ \nu_e \mu^- \bar{\nu}_{\mu} b \bar{b}$ and the complexity of its contributing QCD and EW loops, the stability of its reduction is comparable to that of other processes with a maximum rank of 4.

In an actual Monte-Carlo integration, the phase-space points encountered are not uniformly distributed and as a result the stability profile can potentially be different in this context.
For this reason, we also show the stability profiles obtained by considering the kinematics of unweighted events generated at LO accuracy for the LHC14 collider setup using the NNPDF 2.3 (NLO) PDF set~\cite{Ball:2012cx}. Except for \IREGI, we find no qualitative difference and the Monte-Carlo distributions even tend to be slightly more stable for \Ninja, showing that its reduction is mainly insensitive to a change of reference frame and c.o.m energy.

In appendix~\ref{app:stability}, we show the stability profiles of processes A)-D) for \emph{all} applicable reduction tools. These further establish the observations drawn in this section.

\section{Conclusions}
\label{sec:conclusions}

We presented an algorithm for the generation of the expansion terms
needed by the \emph{one-loop integrand reduction via Laurent
  expansion} implemented in the public library \textsc{Ninja} from the
numerical components of a tensor numerator.  We have shown how, within a
numerical calculation, these expansion terms can be obtained by
contracting the tensor numerator with appropriate cut-dependent
tensors, which in turn can be efficiently built by means of simple
recursive relations.

The algorithm has been implemented in the most recent version of the
public library \textsc{Ninja}, which can thus be used by tensor-based
one-loop generators by providing the numerical entries of the tensor
numerator of an integral.  We interfaced this library to the {\MadLoop}
generator, part of the {\MGaMClong} framework (available from v2.3.4 onward).

This allowed us to extensively study the performance and the numerical
stability of \textsc{Ninja} and compare it with several other
available tools.  In terms of reduction speed, we observe that \Ninja\ outperforms all other reduction tools, and in particular \Samurai\ and \CutTools\ by a factor of about 6 and 3 respectively.
Also, \Ninja's improvement over other tools increases with the complexity of the process.
In terms of reduction stability, \Ninja\ improves on previous integrand reduction techniques by a considerable amount and in general stands on par with tensor integral reduction as implemented in \Golem, which is however more limited and significantly slower.
Our results show that numerical instability with \Ninja\ only becomes problematic, because of the slowdown induced by reprocessing unstable points using quadruple precision arithmetics, for loop numerators of rank 7 and above which are not of immediate concern for current phenomenology at particle colliders. 

The algorithm and the results presented in this paper therefore enhance the
capabilities of \MadLoop\ and the applicability of \Ninja, and will thus be valuable for future high-energy phenomenological studies, 
especially those involving amplitudes featuring loop diagrams characterized by loop numerators of high rank.

\section*{Acknowledgments}
We would like to thank Simon Badger, Pierpaolo Mastrolia, Giovanni Ossola, Stefano Frixione and Fabio Maltoni for useful discussions and comments on the draft, and Goutam Das and Huasheng Shao for their feedback on the use of the interface discussed in this paper.  The work of TP is supported by an STFC Rutherford Grant ST/M004104/1 and the work of VH is supported by the SNF grant PBELP2 146525.

\begin{appendices}
\numberwithin{equation}{section}

\section{Explicit formulas for expansion terms}
\label{app:expl-results-expans}

In this appendix we collect all the formulas for the Laurent expansion
terms defined in sect.~\ref{sec:semi-numer-integr}.  The results
have been obtained as described in sect.~\ref{sec:tens-proj-laur}
and implemented in the \textsc{Ninja} library.  One can check that all
the tensors appearing in the following formulas can be built by means
of the recursive relations in Eq.~\eqref{eq:tenvr},
\eqref{eq:tenv1rmkv2k} and~\eqref{eq:tenv1v2v3}.  It is worth
observing that, for efficiency reasons, in the actual implementation
only a minmal set of terms is computed depending on the cut and on the
rank of the numerator.  We also make sure to reuse contributions to
different terms within the same expansion whenever this is possible.

In the following, we again use the notation introduced in
sect.~\ref{sec:notation}, with the labeling of the terms in each
expansion as defined in sect.~\ref{sec:semi-numer-integr}.
Here is the full list of contributions:
\begin{itemize}
\item evaluation of numerator $\N(q)$
\begin{equation}
  \N(q) = \sum_{r=0}^R \N^{(r)}(q^r),
\end{equation}
\item $\mu^2$-expansion
  \begin{align}
  n_{\mu^2}^{(R)} ={}& \N^{(R)}(v_0^R) \\
  n_{\mu^2}^{(R-1)} ={} & \N^{(R-1)}(v_0^{R-1}) + \N^{(R)}(v_0^{R-1} v_1),
  \end{align}
\item $t_3$-expansion
\begin{align}
  n_{t_3}^{(R,0)} ={}& \N^{(R)}(v_3^R) \\
  n_{t_3}^{(R-1,0)} = {}& \N^{(R-1)}(v_3^{R-1}) + \N^{(R)}(v_3^{R-1} v_0) \\
  n_{t_3}^{(R-2,0)} ={}& \N^{(R-2)}(v_3^{R-2}) + \N^{(R-1)}(v_3^{R-2} v_0)  + \N^{(R)}(v_3^{R-2} v_0^2) + \beta\, \N^{(R)}(v_3^{R-1} v_4) \\
  n_{t_3}^{(R-2,1)} ={}& \N^{(R)}(v_3^{R-1} v_4) \\
  n_{t_3}^{(R-3,0)} ={}& \N^{(R-3)}(v_3^{R-3}) + \N^{(R-2)}(v_3^{R-3} v_0)  + \N^{(R-1)}(v_3^{R-3} v_0^2) + \N^{(R)}(v_3^{R-3} v_0^3) \nn & + \beta\, \Big(\N^{(R-1)}(v_3^{R-2} v_4) + \N^{(R)}(v_3^{R-2} v_0 v_4) \Big) \\
  n_{t_3}^{(R-3,1)} ={}& \N^{(R-1)}(v_3^{R-2} v_4) + \N^{(R)}(v_3^{R-2} v_0 v_4) \\
  n_{t_3}^{(R-4,0)} ={}& \N^{(R-4)}(v_3^{R-4}) + \N^{(R-3)}(v_3^{R-4} v_0) \nn & + \N^{(R-2)}(v_3^{R-4} v_0^2)  + \N^{(R-1)}(v_3^{R-4} v_0^3) + \N^{(R)}(v_3^{R-4} v_0^4) \nn & + \beta\, \Big(\N^{(R-2)}(v_3^{R-3} v_4) + \N^{(R-1)}(v_3^{R-3} v_0 v_4) + \N^{(R)}(v_3^{R-3} v_0^2 v_4) \Big) \nn & + \beta^2\, \N^{(R)}(v_3^{R-2} v_4^2) \\
  n_{t_3}^{(R-4,1)} ={}& \N^{(R-2)}(v_3^{R-3} v_4) + \N^{(R-1)}(v_3^{R-3} v_0 v_4) + \N^{(R)}(v_3^{R-3} v_0^2 v_4) \nn & + 2\, \beta\, \N^{(R)}(v_3^{R-2} v_4^2)  \\
  n_{t_3}^{(R-4,2)} ={}& \N^{(R)}(v_3^{R-2} v_4^2) ,
\end{align}
\item $t_2$-expansion
  \begin{align}
    n_{t_3}^{(R,0,0)} ={}& \N^{(R)}(v_3^R) \\
    n_{t_3}^{(R-1,0,0)} = {}& \N^{(R-1)}(v_3^{R-1}) + \N^{(R)}(v_3^{R-1} v_1) \\
    n_{t_3}^{(R-1,1,0)} = {}& \N^{(R)}(v_3^{R-1} v_2) \\
    n_{t_3}^{(R-2,0,0)} = {}& \N^{(R-2)}(v_3^{R-2}) + \N^{(R-1)}(v_3^{R-2} v_1) + \N^{(R)}(v_3^{R-2} v_1^2) \nn & + \beta_0\, \N^{(R)}(v_3^{R-1} v_4) \\
    n_{t_3}^{(R-2,0,1)} = {}& \N^{(R)}(v_3^{R-1} v_4) \\
    n_{t_3}^{(R-2,1,0)} = {}& \N^{(R-1)}(v_3^{R-2} v_2) + \N^{(R)}(v_3^{R-2} v_1 v_2) + \beta_1\, \N^{(R)}(v_3^{R-1} v_4) \\
    n_{t_3}^{(R-2,2,0)} = {}& \N^{(R)}(v_3^{R-2} v_2^2) + \beta_2\, \N^{(R)}(v_3^{R-1} v_4) \\
    n_{t_3}^{(R-3,0,0)} = {}& \N^{(R-3)}(v_3^{R-3}) + \N^{(R-2)}(v_3^{R-3} v_1) \nn & + \N^{(R-1)}(v_3^{R-3} v_1^2) + \N^{(R)}(v_3^{R-3} v_1^3) \nn & + \beta_0\, \Big(\N^{(R)}(v_3^{R-2} v_1 v_4) + \N^{(R-1)}(v_3^{R-2} v_4) \Big) \\
    n_{t_3}^{(R-3,0,1)} = {}& \N^{(R)}(v_3^{R-2} v_1 v_4) + \N^{(R-1)}(v_3^{R-2} v_4) \\
    n_{t_3}^{(R-3,1,0)} = {}&  \N^{(R-2)}(v_3^{R-3} v_2) + \N^{(R-1)}(v_3^{R-3} v_1 v_2) + \N^{(R)}(v_3^{R-3} v_1^2 v_2)  \nn & + \beta_0\, \N^{(R)}(v_3^{R-2} v_2 v_4) + \beta_1\, \Big(\N^{(R)}(v_3^{R-2} v_1 v_4) + \N^{(R-1)}(v_3^{R-2} v_4) \Big) \\
    n_{t_3}^{(R-3,1,1)} ={}& \N^{(R)}(v_3^{R-2} v_2 v_4) \\
    n_{t_3}^{(R-3,2,0)} = {}&  \N^{(R-1)}(v_3^{R-3} v_2^2) + \N^{(R)}(v_3^{R-3} v_1 v_2^2) + \beta_1\, \N^{(R)}(v_3^{R-2} v_2 v_4) \nn & + \beta_2 \Big(\N^{(R)}(v_3^{R-2} v_1 v_4) + \N^{(R-1)}(v_3^{R-2} v_4) \Big) \\
    n_{t_3}^{(R-3,3,0)} = {}&  \N^{(R)}(v_3^{R-3} v_2^3) + \beta_2\, \N^{(R)}(v_3^{R-2} v_2 v_4).
  \end{align}
\end{itemize}

\newpage

\section{Details of timing performances}
\label{app:timing_table}

This section presents \MadLoop\ timing profile for the generation and the computation of the tensor numerator components for all the benchmark processes A)-D) introduced at the beginning of sect.~\ref{sec:applications}. We also show the time necessary for performing the loop reduction with each of the six reduction tools interfaced to \MadLoop\ (when available): \IREGI, \PJFry, \Golem, \CutTools, \Samurai\ and \Ninja. The reader can easily reproduce analogous results for various compilers and machines by using the automated {\tt check timing} command of the \MGaMC\ interface.


\begin{table}[!htbp]
\begin{tabular}{L{5cm} C{2cm} C{2cm} C{2.2cm} C{2.2cm}}
Timing profile				& $u \bar{u} \rightarrow 5 \cdot Z$	& $u \bar{u} \rightarrow e^+ \nu_e \mu^-\bar{\nu}_{\mu} b \bar{b}$	&
                                                       $u \bar{u} \rightarrow t \bar{t} b \bar{b} d \bar{d}$ 	& $u \bar{u} \rightarrow t \bar{t} b \bar{b} d \bar{d} g$ 							\\
\toprule
Generation time			& 45 min 					& 2h10m 							& 16 min	 					& $\sim$2 days 			\\
\# Loop diagrams			& 3870 					& 51363 							& 5875						& 99981					\\
\# Loop groups				& 1051 					& 4343 							& 1716						& 18469					\\
Maximum numerator rank		& 6 						& 4 								& 4							& 5						\\
Integrand computation time	& 0.17 s 					& 0.88 s 							& 0.12 s						& 24 s					\\
\midrule
\Ninja\ reduction time		& 0.14 s 					& 0.42 s 							& 0.10 s						& 2.0 s					\\
\Samurai\ reduction time		& 0.96 s					& 2.7 s 							& 0.58 s						& 18.1 s					\\
\CutTools\ reduction time		& 0.58 s 					& 1.5 s 							& 0.27 s						& 7.3 s					\\
\midrule[\heavyrulewidth]
						& $g g \rightarrow 2 \cdot Z$		& $g g \rightarrow 3 \cdot Z$				& $g g \rightarrow 4 \cdot Z $		& $g g \rightarrow 5 \cdot Z$	\\
\midrule[\heavyrulewidth]
Generation time			& 15.6 s					& 110 s							& 26 min						& 29 h				\\
\# Loop diagrams			& 28						& 156							& 1176 						&11700				\\
\# Loop groups				& 28						& 156							& 1176 						&11700				\\
Maximum numerator rank		& 4						& 5								& 6 							& 7					\\
Integrand computation time	& 0.95 ms					& 11 ms							& 0.17 s 						& 2.5 s				\\
\midrule
\Ninja\ reduction time		& 1.5 ms					& 20 ms							& 0.32 s 						& 5.5 s				\\
\Samurai\ reduction time		& 6.3 ms					& 90 ms							& 1.8 s 						& 39 s				\\
\CutTools\ reduction time		& 4.0 ms					& 58 ms							& 1.1 s 						& 21 s				\\
\midrule
\IREGI\ reduction time		& 31 ms 					& 1.9 s							& 120 s 						& \NA				\\
\Golem\ reduction time		& 20 ms					& 0.94 s							& 28 s 						& \NA				\\
\PJFry\ reduction time		& 5 ms					& 0.36 s							& \NA						& \NA				\\
\bottomrule
\end{tabular}
\caption{\label{tab:timing_high_mult} The upper table presents results for processes with high-multiplicity and low loop numerator ranks. Notice that the process $u \bar{u} \rightarrow e^+ \nu_e \mu^- \bar{\nu}_{\mu} b \bar{b}$ (massless $b$-quarks) includes \emph{all} SM tree and loop contributions (\emph{i.e.} of both QCD and EW origin, resonant as well as non-resonant ones and also including contributions of order $\mathcal{O}(\alpha_s^0)$). The timing of the loop matrix element $2\Re(\mathcal{A}^{(loop)}\mathcal{A}^{(tree)\dagger})$ of the process $u \bar{u} \rightarrow Z Z Z Z Z$ (denoted $u \bar{u} \rightarrow 5 \cdot Z$ in the table) echoes the profiling presented in the lower table for the evaluation of the loop-induced matrix element $|\mathcal{A}^{(loop)}|^2$ of the gluon fusion contributions up to the same final states.
All timings in this table refer to the the computation of the loop matrix element summed over colors but for a \emph{single} helicity configuration. The test machine is using a single core (for process generation as well) of an Intel Core i7 CPU (2.7 GHz) and the executable is compiled with GNU {\tt gfortran -O2} (v4.8.2).}
\end{table}


\begin{table}[ph!]
\begin{tabular}{L{5cm} C{2cm} C{2cm} C{2.2cm} C{2.2cm}}
Timing profile				& $g g \rightarrow t \bar{t}$	& $g g \rightarrow t \bar{t} g$	& $g g \rightarrow t \bar{t} g g$	& $g g \rightarrow t \bar{t} g g g$ 	\\
\toprule
Generation time			& 14 s					& 46 s					& 14 min					& $\sim$2 days 				\\
\# Loop diagrams			& 36						& 427					& 5547					& 82470						\\
\# Loop groups				& 26						& 164					& 1168					& 9940						\\
Maximum numerator rank		& 3						& 4						& 5						& 6 							\\
Integrand computation time	& 0.34 ms					& 5.5 ms					& 215 ms					& 36 s 						\\
\midrule
\Ninja\ reduction time		& 0.38 ms					& 5.7 ms					& 85 ms					& 0.78 s						\\
\Samurai\ reduction time		& 1.9 ms					& 22 ms					& 362 ms					& 9.5 s						\\
\CutTools\ reduction time		& 1.0 ms					& 14 ms					& 240 ms					& 5.4 s						\\
\midrule
\IREGI\ reduction time		& 5.5 ms					& 290 ms					& 12.8 s					& 13 min						\\
\Golem\ reduction time		& 4.6 ms					& 113 ms					& 3.4 s					& \NA						\\
\PJFry\ reduction time		& 1.1 ms					& 50 ms					& \NA					& \NA						\\
\midrule[\heavyrulewidth]
						& $g g \rightarrow H g$		& $g g \rightarrow H g g$		& $g g \rightarrow H g g g$	&\\
\midrule[\heavyrulewidth]
Generation time			& 28 s					& 122 s					& 1h50m					&\\
\# Loop diagrams			& 69						& 875					& 12300					&\\
\# Loop groups				& 22						& 122					& 836					&\\
Maximum numerator rank		& 5						& 6						& 7						&\\
Integrand computation time	& 0.96 ms					& 19.2 ms					& 1.1 s					&\\
\midrule
\Ninja\ reduction time		& 0.55 ms					& 8.7 ms					& 130 ms					&\\
\Samurai\ reduction time		& 3.2 ms					& 47 ms					& 876 ms					&\\
\CutTools\ reduction time		& 1.9 ms					& 32 ms					& 580 ms					&\\
\midrule
\IREGI\ reduction time		& 18.6 ms					& 1.7 s					& \NA					&\\
\Golem\ reduction time		& 6.9 ms					& 0.28 s					& \NA					&\\
\midrule[\heavyrulewidth]
						& $g g \rightarrow Y g$		& $g g \rightarrow Y g g$		& $g g \rightarrow Y g g g$	&\\
\midrule[\heavyrulewidth]
Generation time			& 72 s					& 234 s					& 3h20m 					&\\
\# Loop diagrams			& 141					& 1463					& 18420					&\\
\# Loop groups				& 31						& 163					& 1111					&\\
Maximum numerator rank		& 5						& 6 						& 7						&\\
Integrand computation time	& 3.0 ms					& 47 ms 					& 1.8 s					&\\
\midrule
\Ninja\ reduction time		& 1.4 ms					& 20 ms					& 0.27 s					&\\
\Samurai\ reduction time		& 5.8 ms					& 85 ms					& 1.7 s					&\\
\midrule
\IREGI\ reduction time		& 35.4 ms					& 3.5 s					& \NA 					&\\
\Golem\ reduction time		& 12.5 ms					& 0.50 s					& \NA 					&\\
\bottomrule
\end{tabular}
\caption{\label{tab:timing_gg_Bng} Same setup as described in the caption of table~\ref{tab:timing_high_mult}. Profiling of the runtime of the processes $g g \rightarrow \{X\} + n\cdot g$ with $\{X\}=t\bar{t},H,Y$ and $n=(0,)1,2,3$. The symbol $Y$ denotes a spin-2 particle with a mass of 1 TeV and interactions as described in sect. 2.3 of ref.~\cite{Artoisenet:2013puc}.}
\end{table}


The numbers shown in tables~\ref{tab:timing_high_mult} and \ref{tab:timing_gg_Bng} refer to the computation of the loop matrix element averaged and summed over color assignments but \emph{for a single helicity} and kinematic configuration. This is the relevant figure of merit in \MGaMC\ since it implements a Monte-Carlo over helicity (with importance sampling) when integrating loop contributions. It is worth noting however that for loop matrix elements with a Born contribution, only the numerator computation time ($t_{\textrm{num}}$) scales with the number of contributing (analytically non-zero) helicity configurations ($n_{\textrm{hel}}$) whereas the reduction time ($t_{\textrm{red}}$) remains constant as the integrand numerators can be summed over helicity configurations before being reduced. The total time for the computation of the loop matrix element summed over all helicity configurations can therefore simply be computed as 
\begin{equation}
t_{\textrm{tot, hel+color summed}} = n_{\textrm{hel}}\cdot t_{\textrm{num}} + t_{\textrm{red}}
\end{equation}
since \MadLoop\ does not implement optimizations across different helicity configurations.

For loop-induced processes however~\cite{Hirschi:2015iia}, this is not possible when using reduction at the integrand level, in which case both $t_{\textrm{num}}$ and $t_{\textrm{red}}$ scale with $n_{\textrm{hel}}$.

The number of loop diagrams indicated does not count the multiple copies with different quark flavors in the loop. A loop group refers to a group of loops which can be reduced together. This number is equal to the number of loops for loop-induced processes since each loop must be reduced individually in this case; otherwise it regroups all loops sharing the same topology (ordered list of denominators identified by their mass $m_i$ and four-momentum flows $p_i$).
 
The synthetic fig.~\ref{fig:TimingPlot} of the main text illustrates best the results and we find that \Ninja\ outperforms all reduction tools considered for all the benchmark processes. We note however that the advantageous apparent exponential growth of the integrand reduction time with the process multiplicity is mitigated by the factorial growth of the time spent in computing the numerator tensor components. This is intrinsic to \MadLoop's approach based on Feynman diagrams which offers maximal flexibility at the expense of not taking advantage of the optimal scaling behavior of recursion relations~\cite{Actis:2012qn}. We stress that \MadLoop\ implements a caching system for recycling part of trees and loops shared across different diagrams. This emulates what recursion relations achieve, but only to a lesser extent even though it already considerably improves the computation time of the tensor numerator.

Generation time is usually not considered as relevant given that it must be performed once per process only. In practice however, this can be of concern since it is typically not easily parallelizable and is also a general hinderance when it comes to testing, debugging or quickly exploring the impact of some modifications to a model. In \MadLoop's approach, generation time is hardly an issue for current phenomenologically relevant processes, but its growth is such that we reached the limit of reasonable process generation for $g g \rightarrow t \bar{t} g g g$ which requires about 2 days of sequential runtime and 40 GB of RAM. 

\newpage

\section{Details of stability performances}
\label{app:stability}

In this appendix, we present the stability profiles obtained for processes A)-D) described at the beginning of sect.~\ref{sec:applications} and for the six loop reductions tools interfaced to \MadLoop\ (when available): \IREGI, \PJFry, \Golem, \CutTools, \Samurai\ and \Ninja. The reader can easily reproduce these profiles by using the automated {\tt check stability} command of the \MGaMC\ interface.

\begin{figure}[ph!]
	\centering
	\includegraphics[trim=80 280 45 70,clip,scale=0.43]{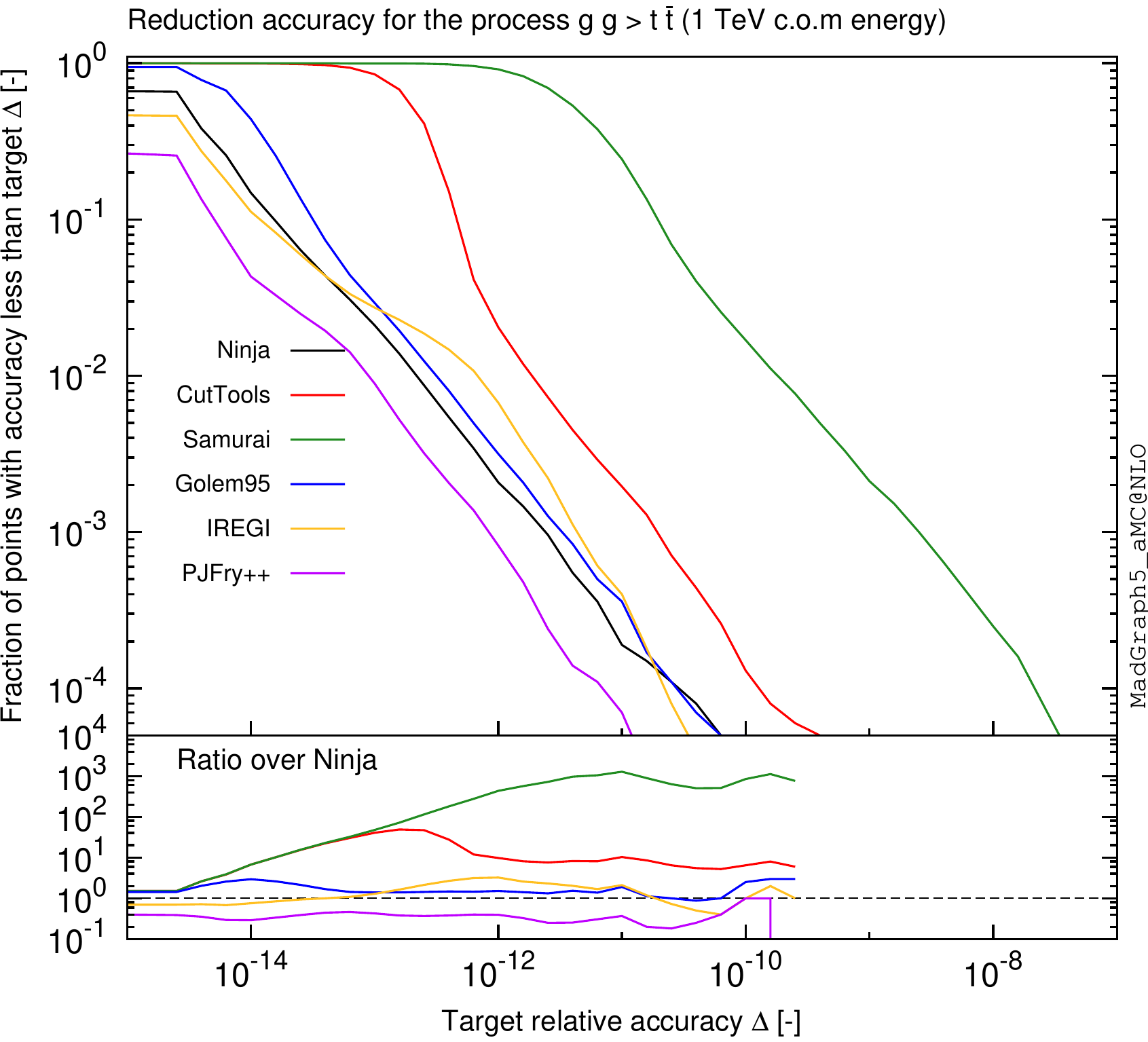}
	\includegraphics[trim=80 280 45 70,clip,scale=0.43]{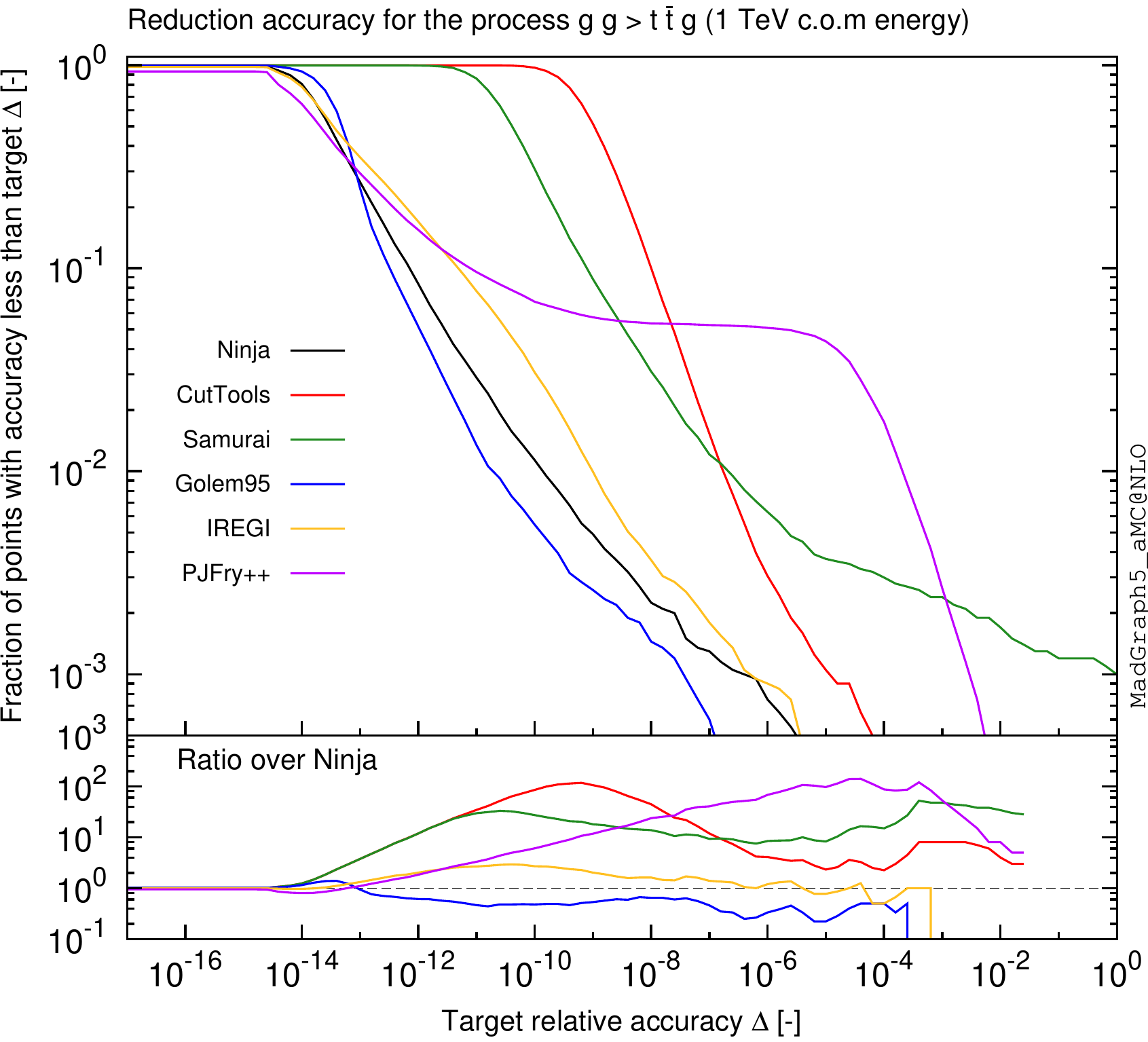}
	\includegraphics[trim=80 280 45 70,clip,scale=0.43]{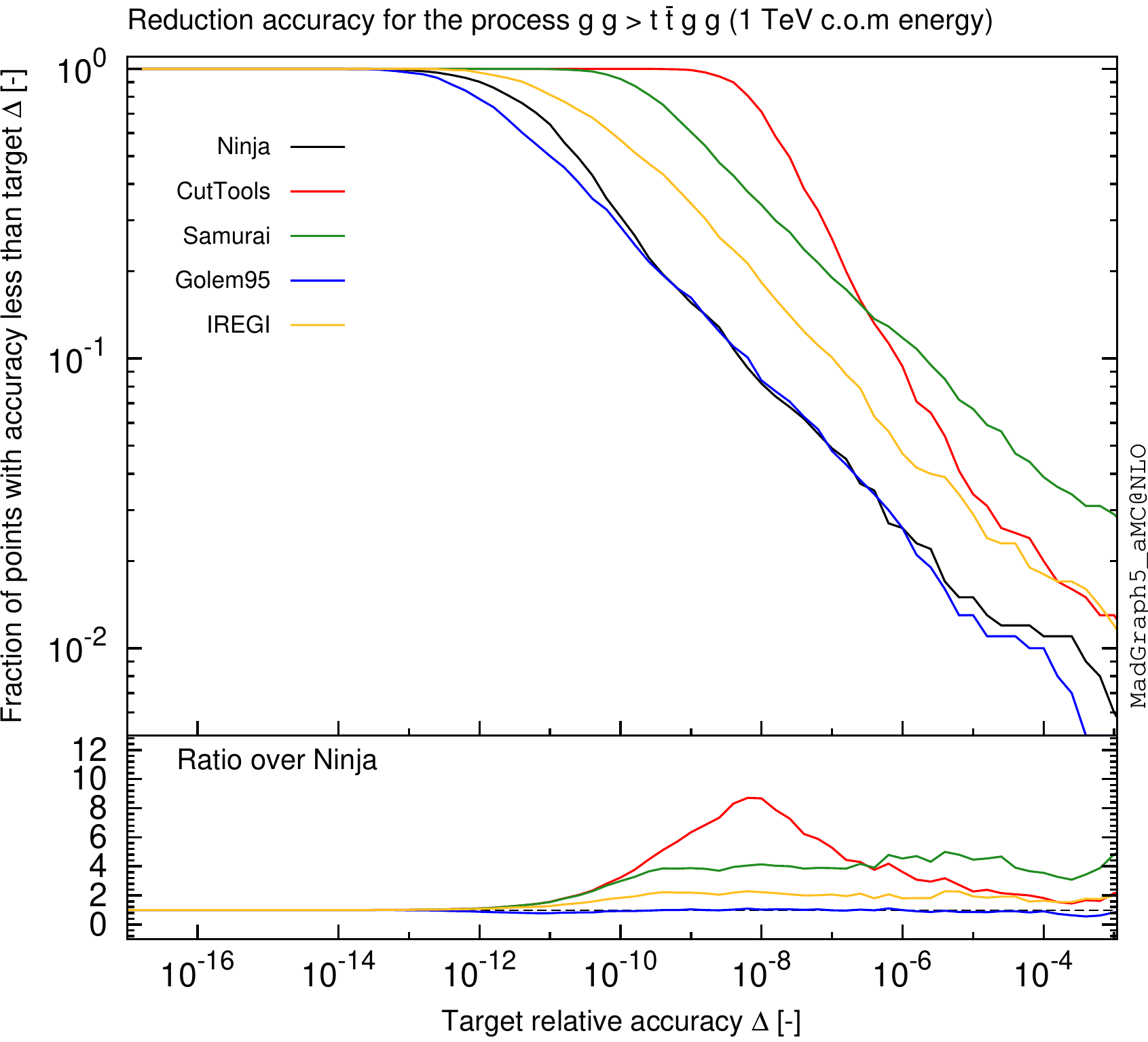}
	\includegraphics[trim=80 280 45 70,clip,scale=0.43]{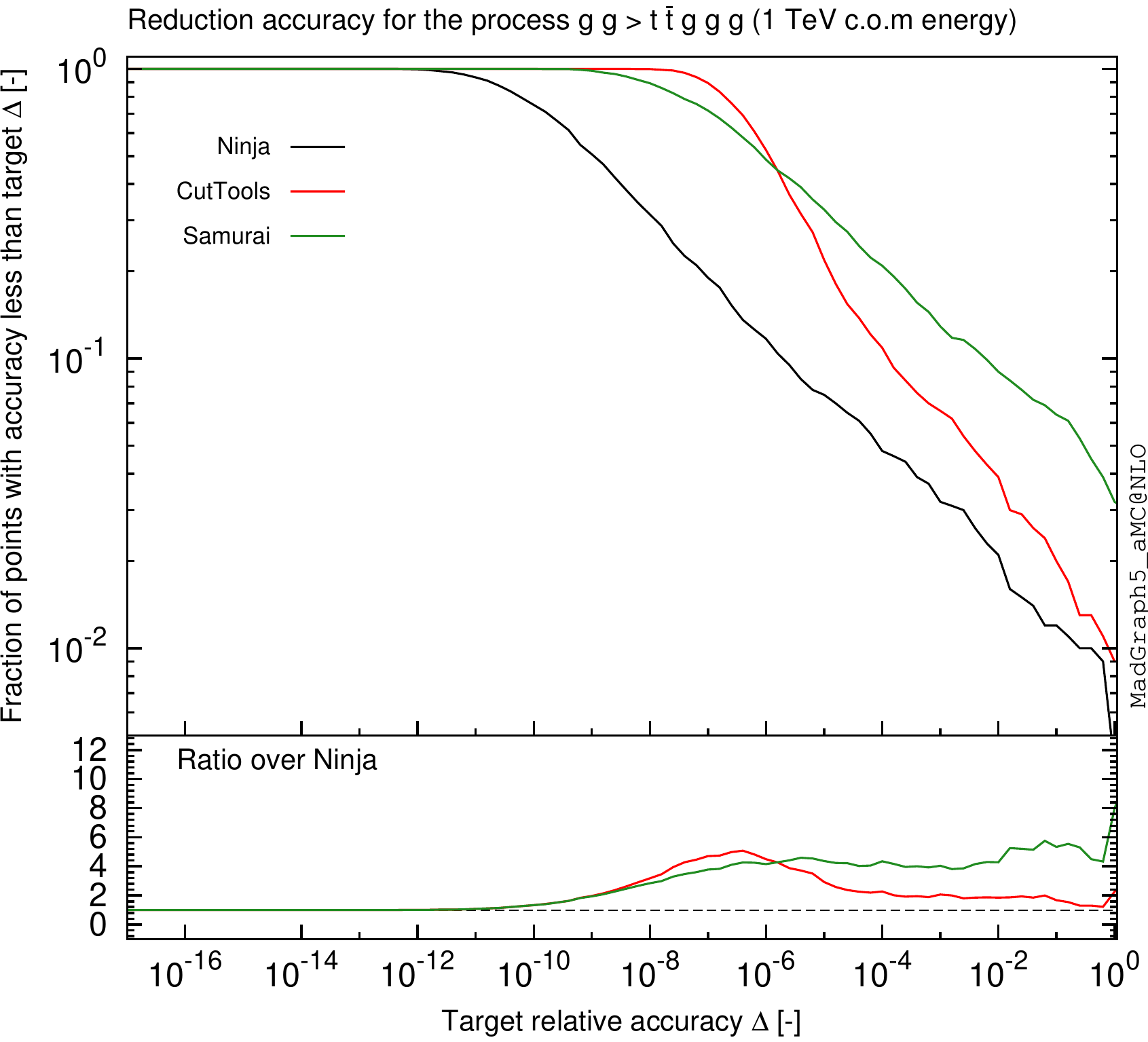}
	\caption{\label{fig:StabilityPlot_ggttx} Same setup as in fig.~\ref{fig:StabilityPlot}, showing the stability profile of all applicable reduction methods interfaced to \MadLoop\ for the class of processes $g g \rightarrow t \bar{t} + \{0,1,2,3\}\cdot g$.}
\end{figure}

Fig.~\ref{fig:StabilityPlot_ggttx} shows results for the process $g g \rightarrow t \bar{t} + \{0,1,2,3\}g$. The plateau displayed by \PJFry\ in $g g \rightarrow t \bar{t} g$ is reminiscent of a known issue in this tool which is now no longer maintained.

\begin{figure}[ph!]
	\centering
	\includegraphics[trim=80 280 45 70,clip,scale=0.43]{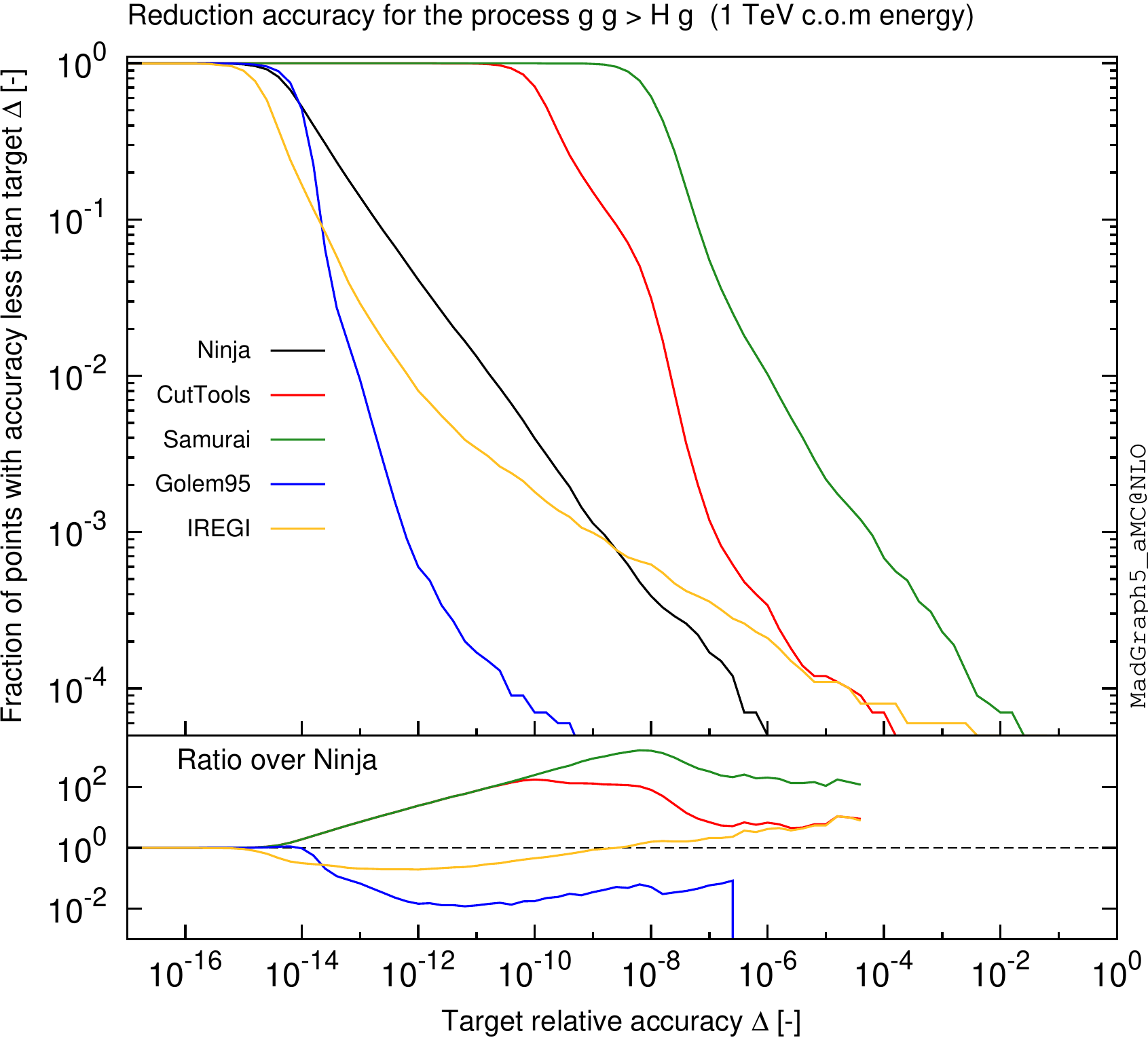}
	\includegraphics[trim=80 280 45 70,clip,scale=0.43]{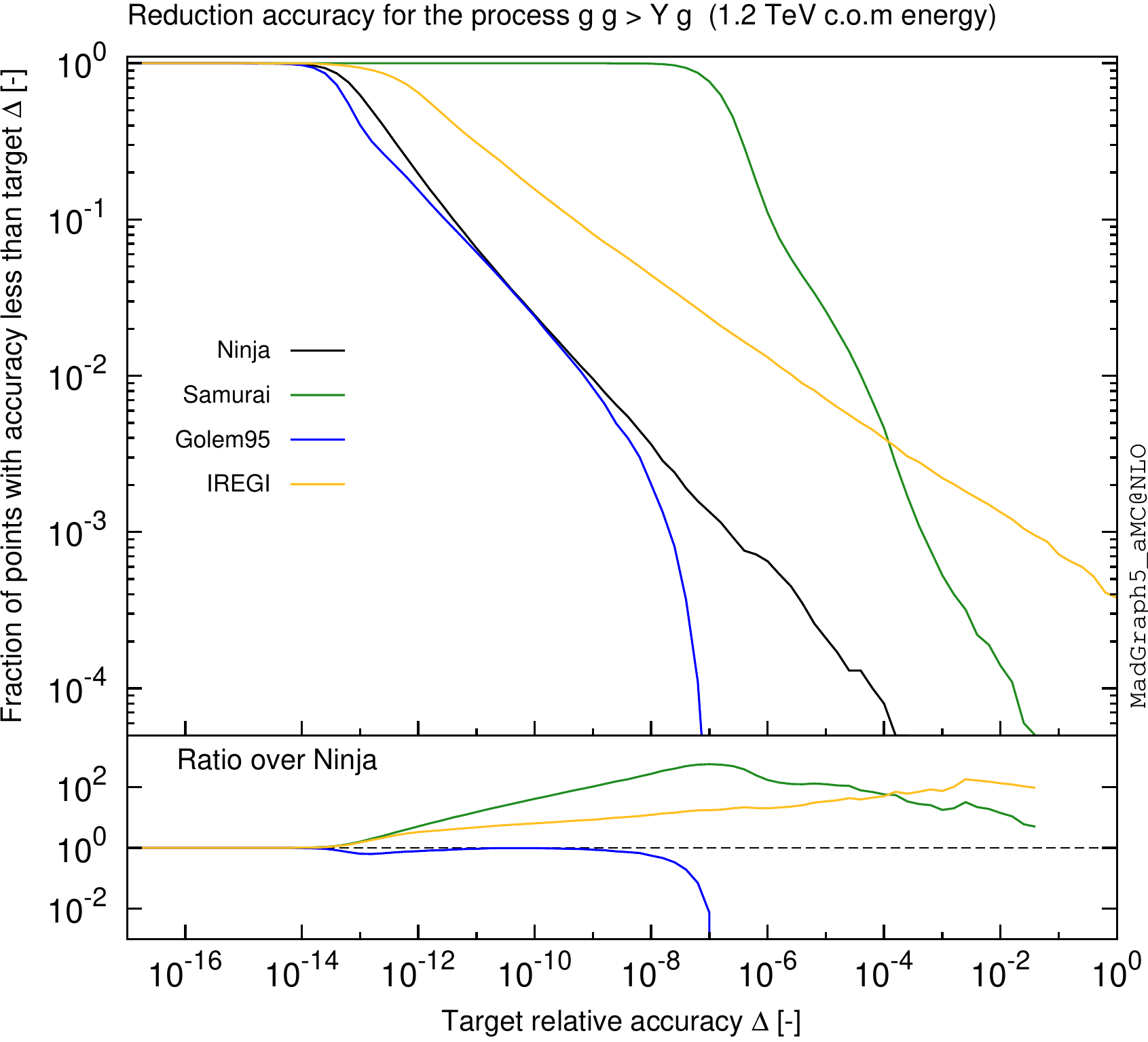}
	\includegraphics[trim=80 280 45 70,clip,scale=0.43]{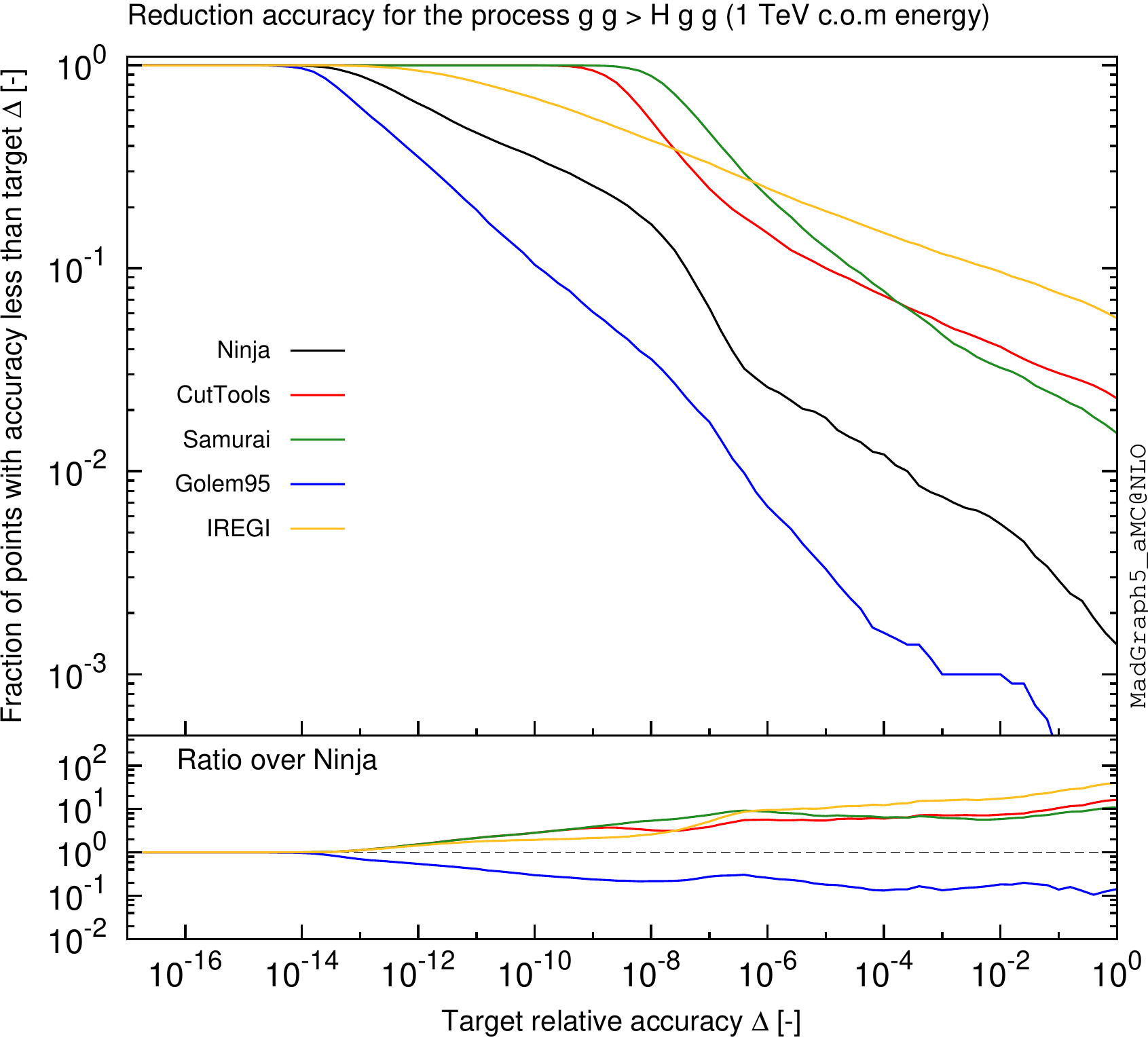}
	\includegraphics[trim=80 280 45 70,clip,scale=0.43]{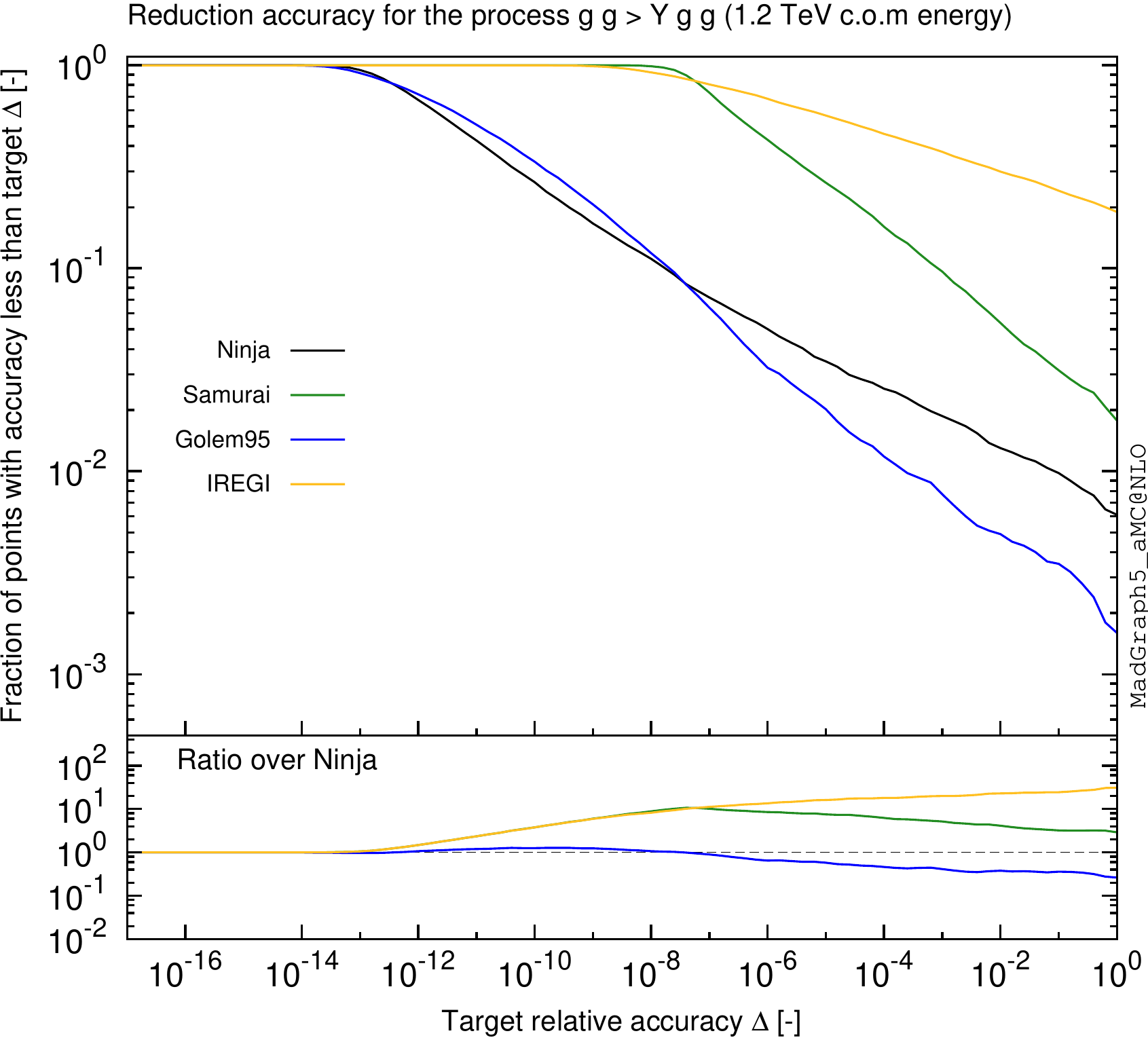}
	\includegraphics[trim=80 280 45 70,clip,scale=0.43]{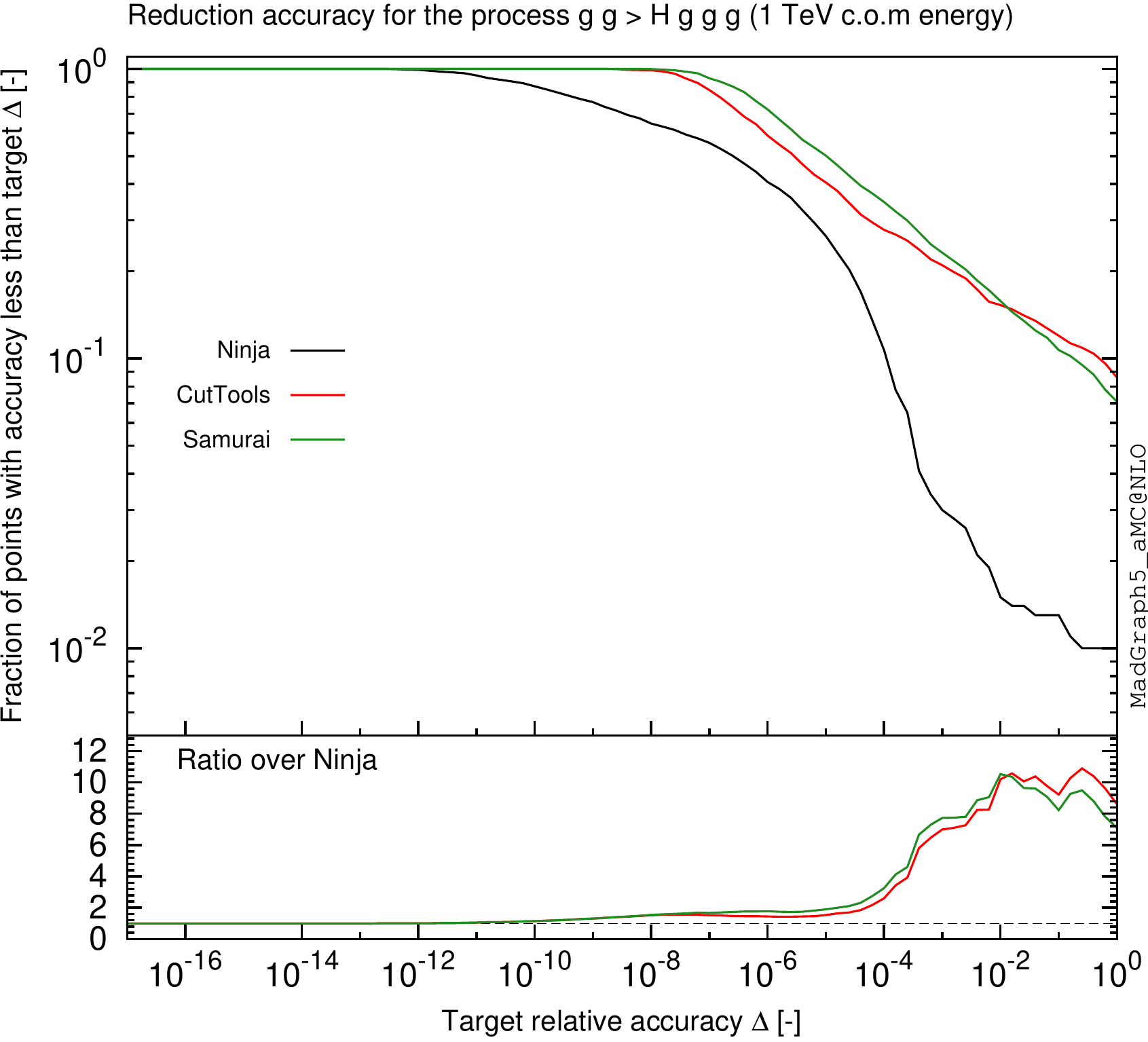}
	\includegraphics[trim=80 280 45 70,clip,scale=0.43]{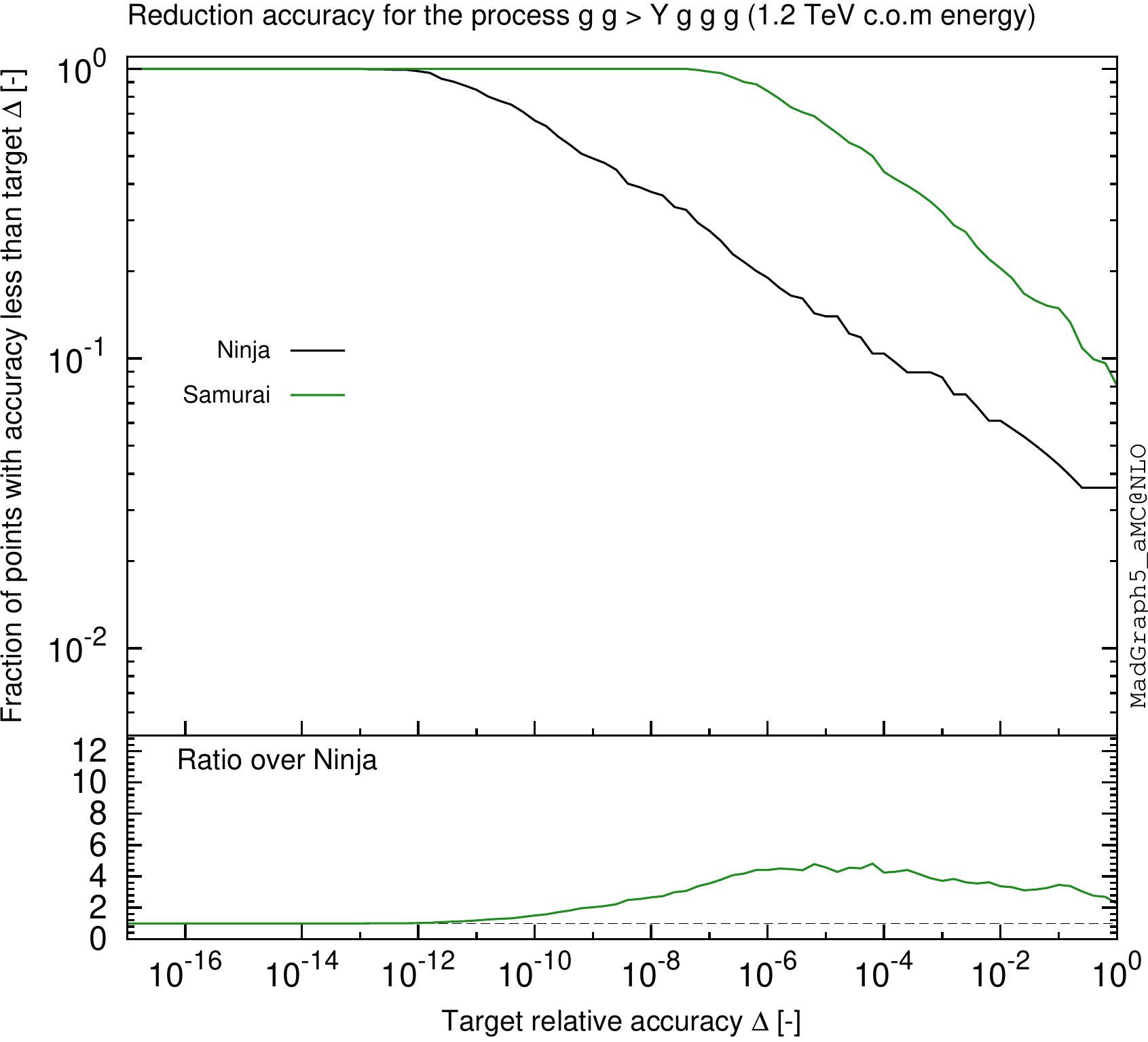}
	\caption{\label{fig:StabilityPlot_y2} Same setup as in fig.~\ref{fig:StabilityPlot}. The plots on the left hand side study processes involving a Higgs in the Higgs Effective Theory model described in ref.~\cite{Artoisenet:2013puc}. The plots on the right hand side study processes involving a spin-2 particle denoted Y with a mass of 1 TeV and interactions as described in sect.~2.3 of ref.~\cite{Artoisenet:2013puc}.}
\end{figure}

Fig.~\ref{fig:StabilityPlot_y2} shows results obtained for the processes $g g \rightarrow H / Y + \{1,2,3\}\cdot g$ involving effective interactions yielding loops of numerator ranks equal to one plus the number of loop denominators. Such loops are particularly challenging to reduce and we observe that the accuracy deteriorates significantly when assuming a completely general form of the higher rank tensor, as it is the case in $g g \rightarrow Y + n\cdot g$, compared to the simpler tensor numerators obtained in $g g \rightarrow H + n\cdot g$.  Indeed, the residues and consequently the fitting procedure are considerably more involved for higher rank integrands.  It is worth noticing that for the simpler tensor structure of $g g \rightarrow H + n\cdot g$ however, the Laurent expansion method is significantly more accurate than other integrand reduction tools, since the numerator expansion methods will return zeroes for vanishing higher rank coefficients, thus avoiding the (inexact) numerical reconstruction of such zeroes from multiple evaluations of the integrand.  In the Higgs case, the stability profile exhibits an unusual shape with a very steep dependence with $\Delta$ in the region $[10^{-5},10^{-3}]$, indicating that any decrease of the default Monte-Carlo stability threshold of $\Delta=10^{-3}$ would have a significant impact on runtime performances as the fraction of points that must be reprocessed using quadruple arithmetics increases.
The comparison of the two profiles obtained using tensor integral reduction highlights the importance of the internal numerical stabilization mechanisms of {\Golem} as the rank of the loop numerator increases.

\begin{figure}[ph!]
	\centering
	\includegraphics[trim=80 280 45 70,clip,scale=0.43]{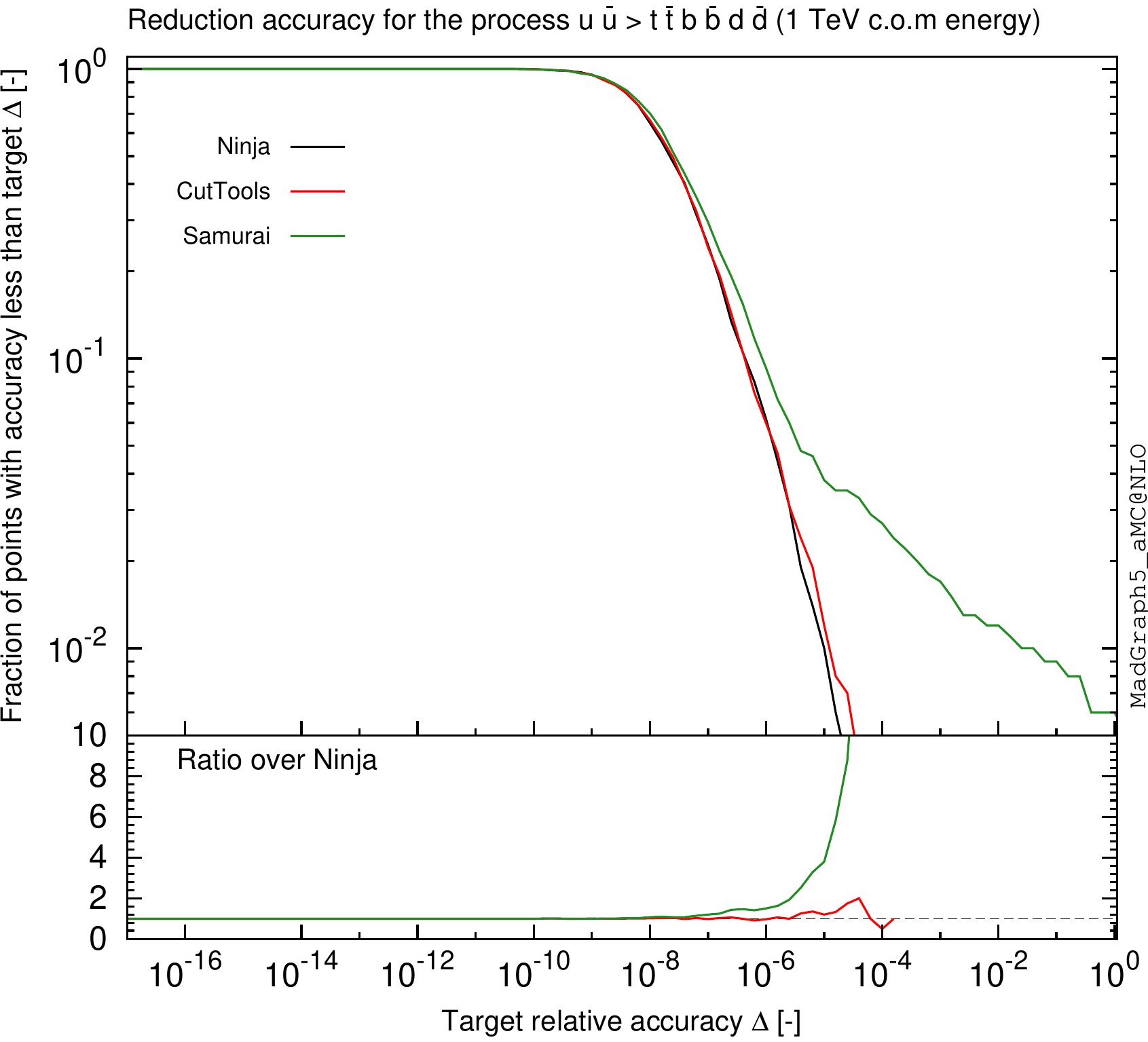}
	\includegraphics[trim=80 280 45 70,clip,scale=0.43]{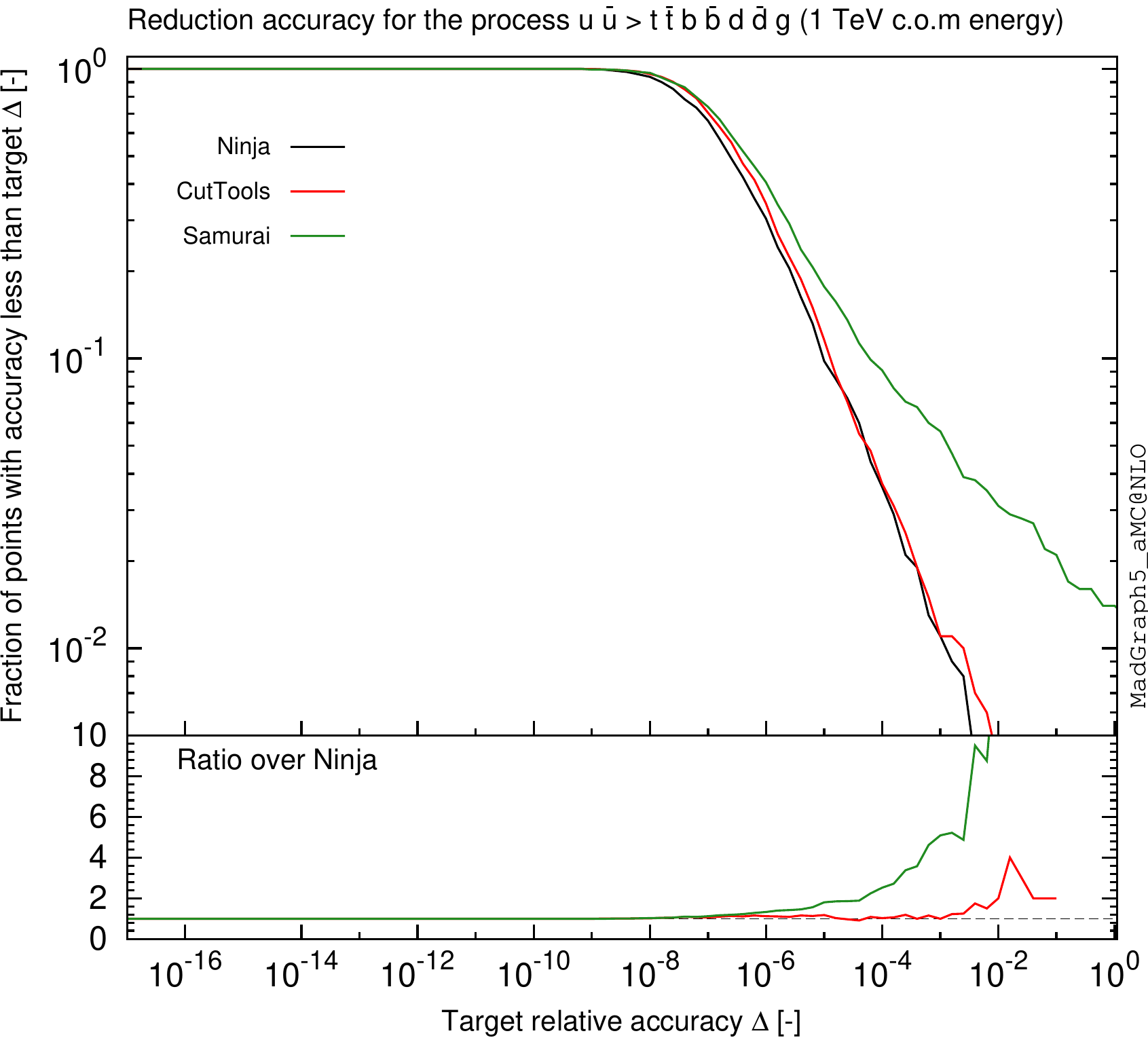}
	\includegraphics[trim=80 280 45 70,clip,scale=0.43]{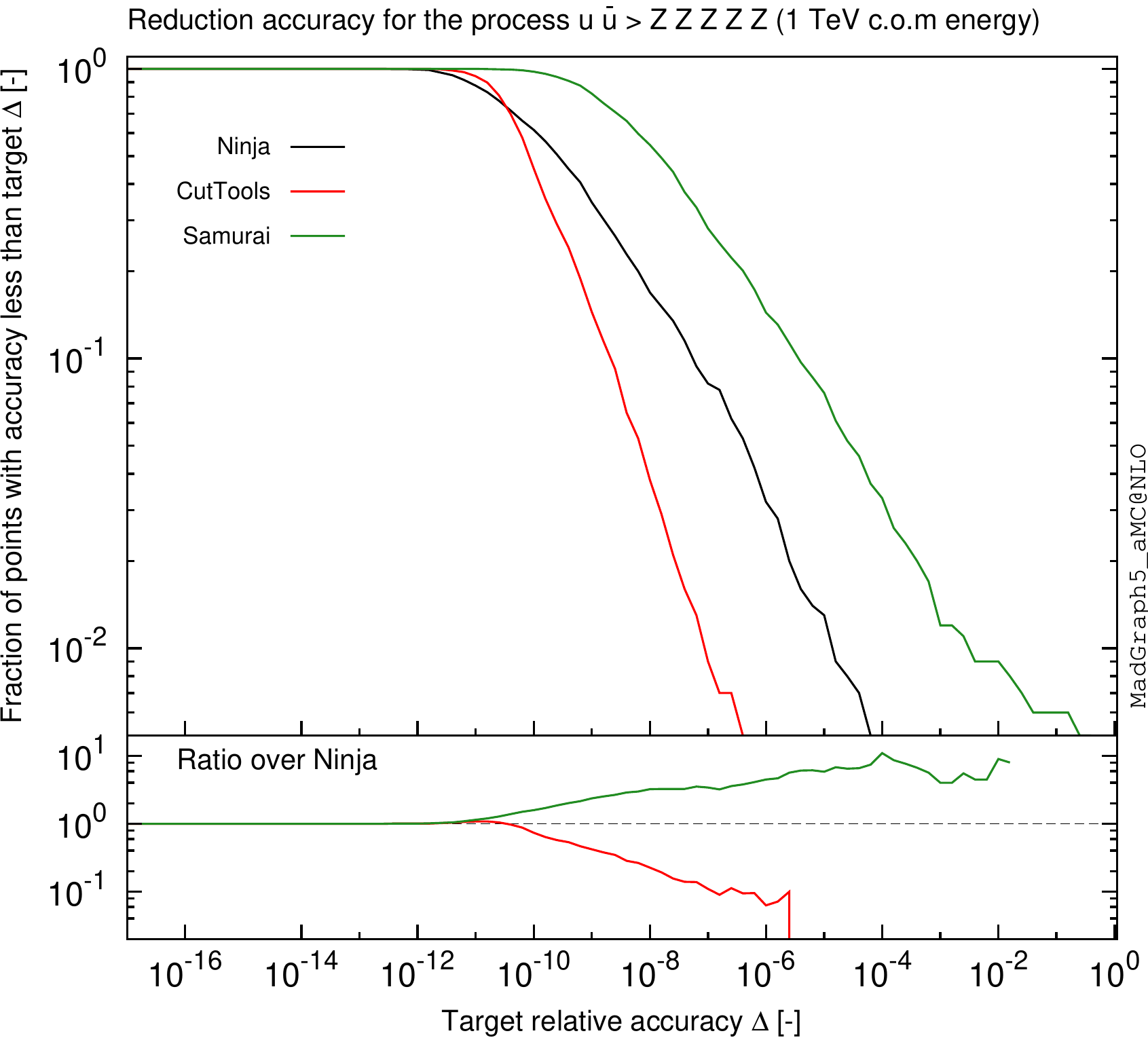}
	\includegraphics[trim=80 280 45 70,clip,scale=0.43]{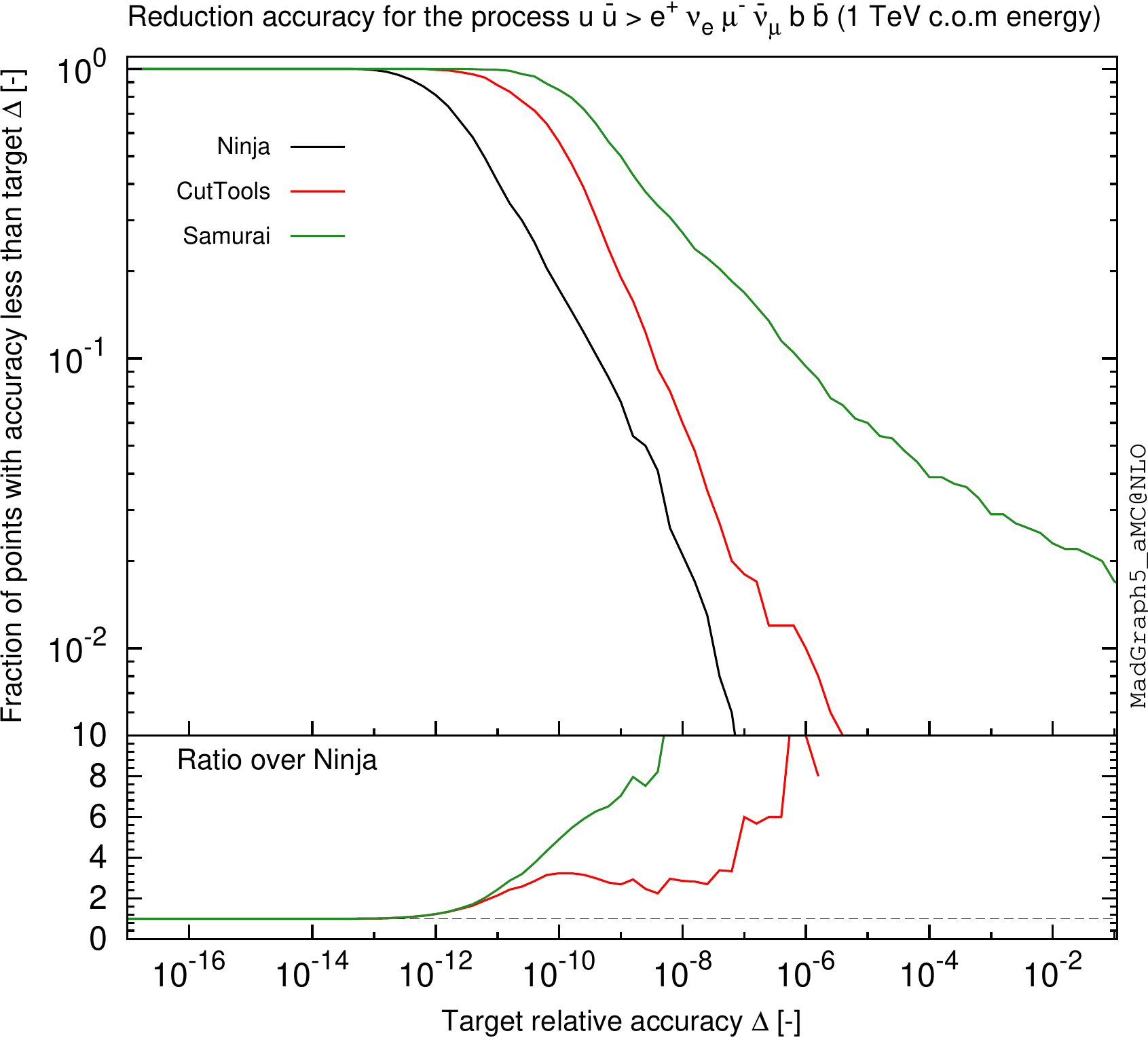}
	\caption{\label{fig:StabilityPlot_8mult} Same setup as in fig.~\ref{fig:StabilityPlot} but for two processes with 6 or more final states. 
	              Notice that the process $u \bar{u} \rightarrow e^+ \nu_e \mu^- \bar{\nu}_{\mu} b \bar{b}$ with massless $b$-quarks (right panel) includes \emph{all} SM tree and loop contributions (i.e.\ of both QCD and EW origin, resonant as well as non-resonant ones and also including contributions of order $\mathcal{O}(\alpha_s^0)$).}
\end{figure}
Fig.~\ref{fig:StabilityPlot_8mult} shows stability results for processes with high multiplicity but relatively low rank. We observe that even though the processes $u \bar{u} \rightarrow t \bar{t} b \bar{b} d \bar{d} (g)$ involve up to 8-(9-)loops, their numerical stability is only slightly worse than that of the lower multiplicity processes with equal maximum rank, such as $g g \rightarrow t \bar{t} g (g)$.
It is interesting to note that when the multiplicity is larger than the rank by several units, the numerical stability of \CutTools\ and \Ninja\ is almost identical, as expected from the fact that the integrand reduction via Laurent expansion method and the traditional OPP numerator fitting are very similar in this limit.  Indeed, when the multiplicity $n$ of the loop lines and the rank $r$ of the numerator satisfy $r\leq n-4$, one can easily show that the result is only determined by the cut-constructible coefficients of the boxes.  These in turn, given their simplicity, are the only coefficients that \textsc{Ninja} computes with the same algorithm as traditional integrand reduction.

The two processes of the bottom insets of fig.~\ref{fig:StabilityPlot_8mult} introduce new scales in the loop amplitudes; first with external massive lines for $u \bar{u} \rightarrow 5\cdot Z$ for which \CutTools\ is slightly more stable than \Ninja\ (unlike for all other processes) and secondly with internal (complex) massive lines for the complete QCD+EW loop contributions to the process $u \bar{u} \rightarrow e^+ \nu_e \mu^- \bar{\nu}_{\mu} b \bar{b}$ for which \Ninja\ is more stable.
\begin{figure}[ph!]
	\centering
	\includegraphics[trim=80 280 45 70,clip,scale=0.43]{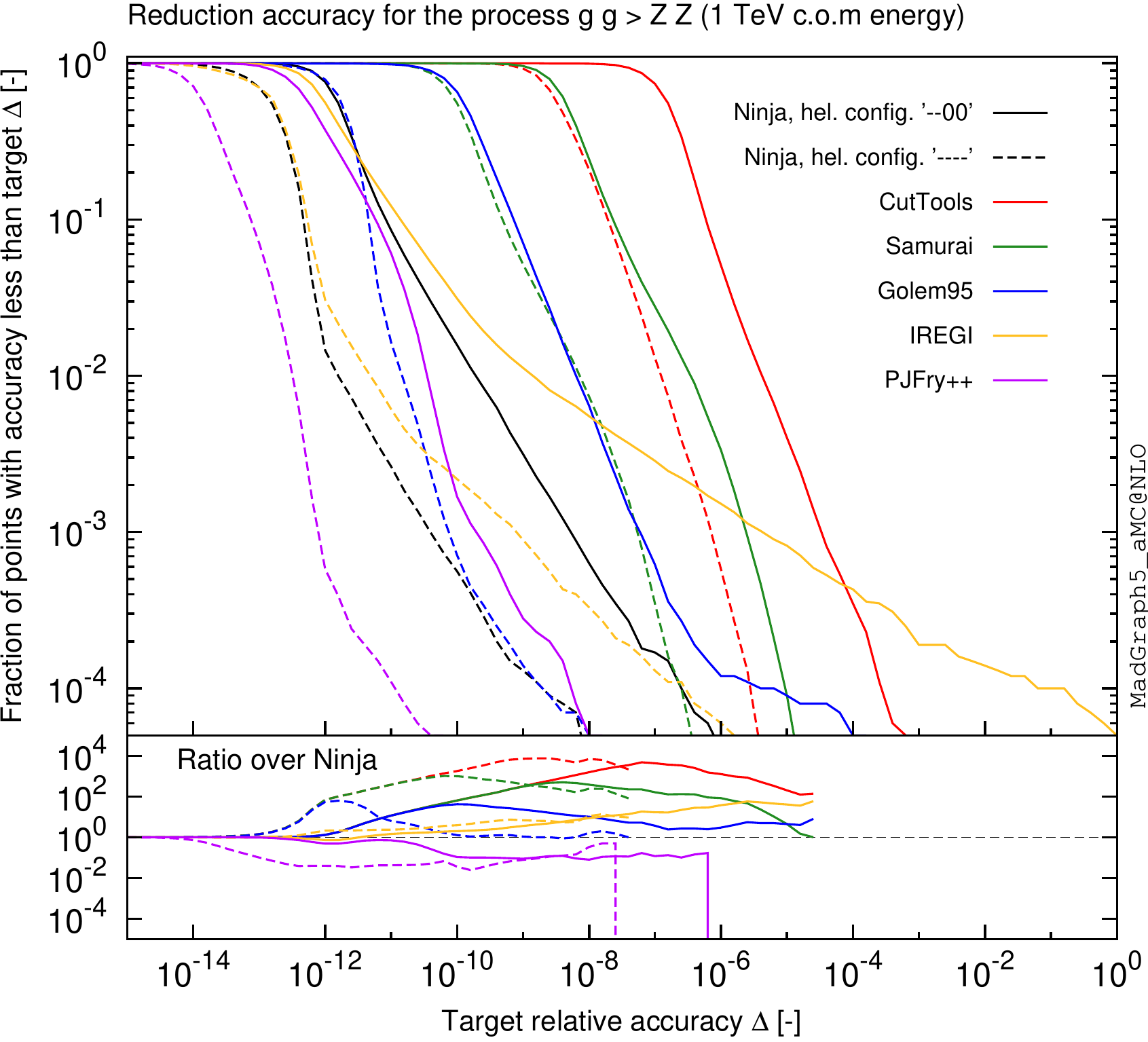}
	\includegraphics[trim=80 280 45 70,clip,scale=0.43]{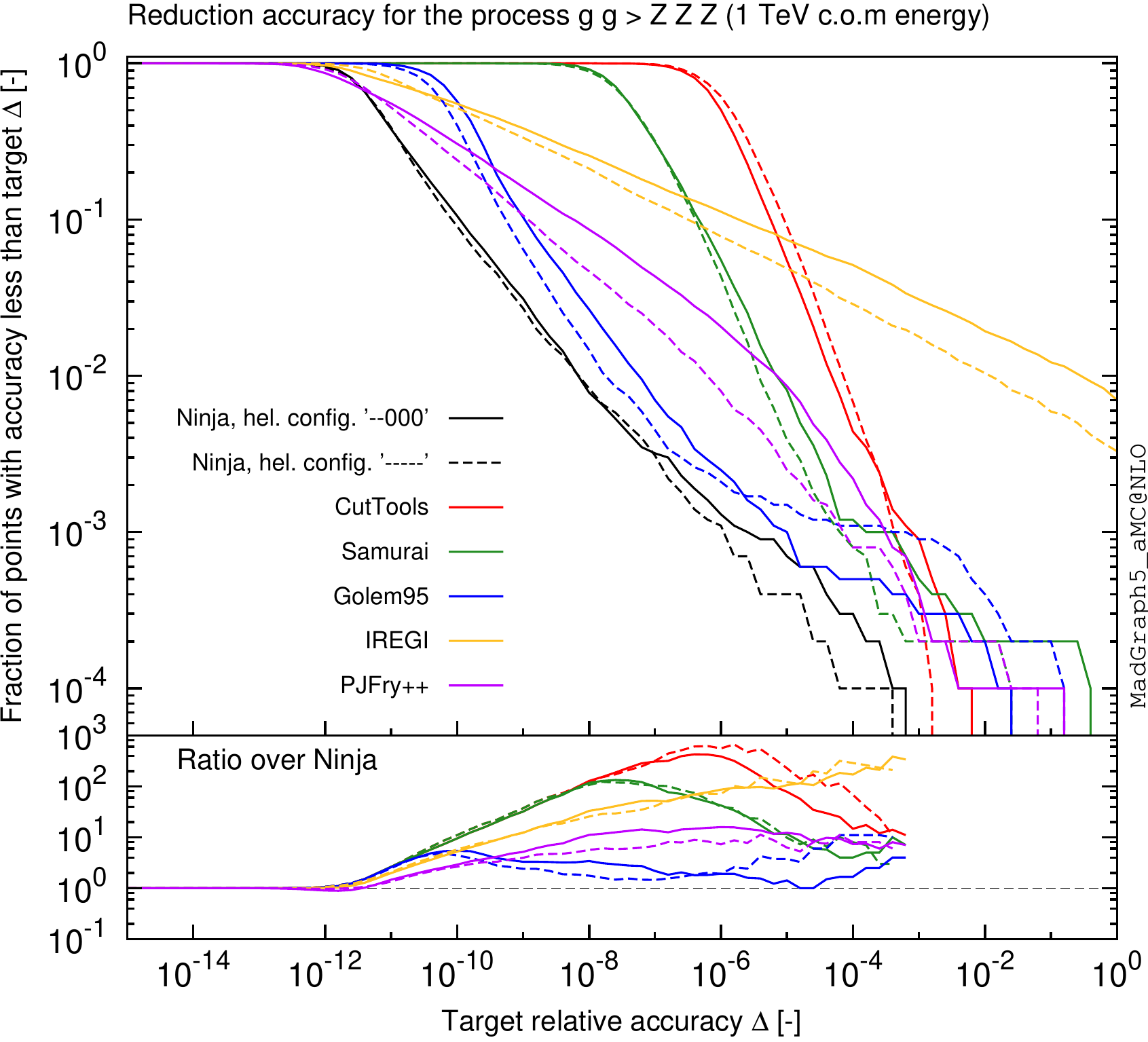}
	\includegraphics[trim=80 280 45 70,clip,scale=0.43]{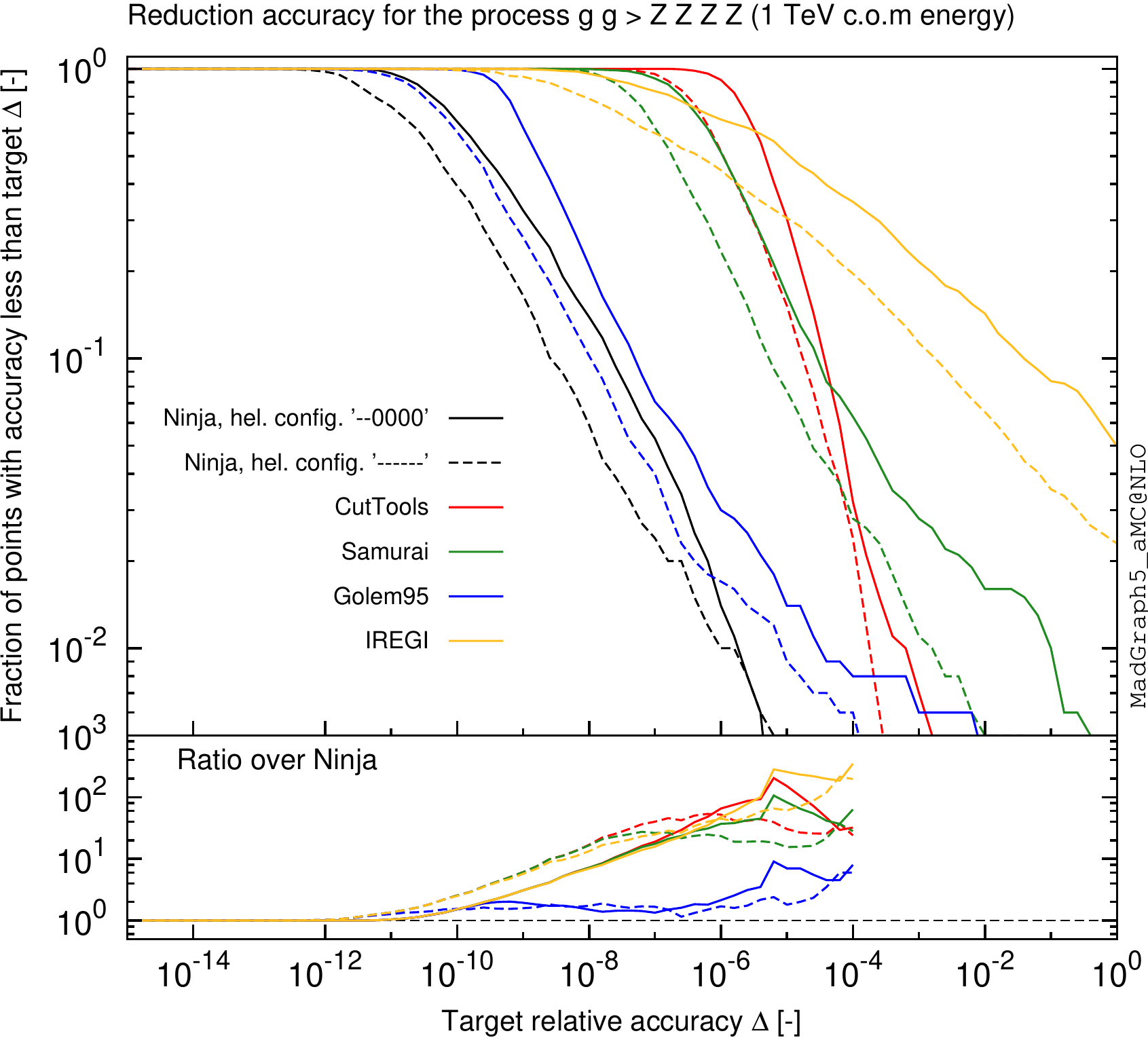}
	\includegraphics[trim=80 280 45 70,clip,scale=0.43]{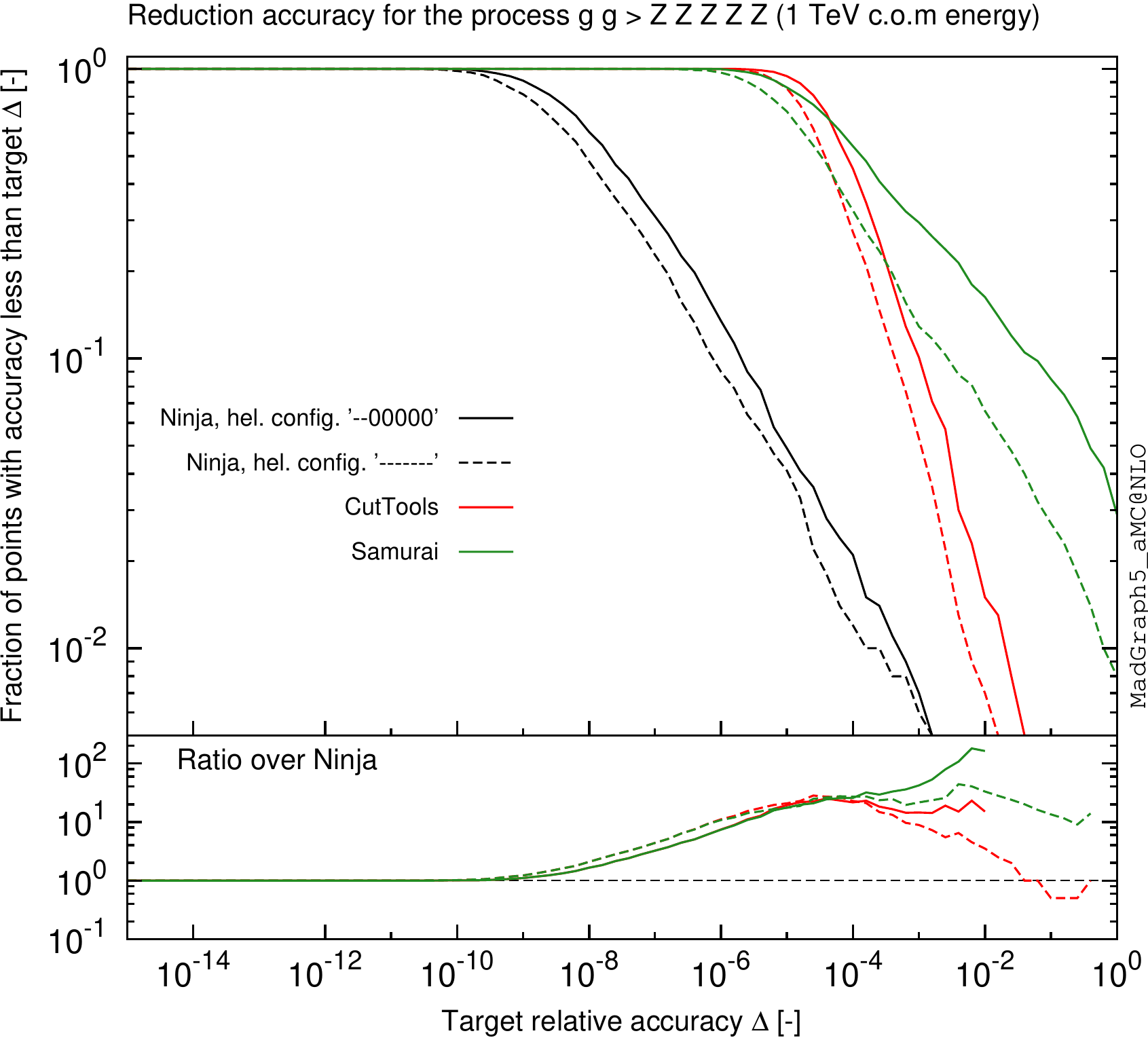}
	\caption{\label{fig:StabilityPlot_zz} Same setup as in fig.~\ref{fig:StabilityPlot} except that no constraint on the final state transverse momenta was applied is applied in this case. The processes considered are loop-induced gluon fusion with up to five $Z$-bosons in the final states. Results are shown for the stability obtained when considering only a single helicity configuration where the two initial state gluons have negative helicity and the final state $Z$-bosons have either helicity 0 (solid line) or a negative one (dashed line).}
\end{figure}
From the numerical stability standpoint, the interesting characteristics of loop-induced processes such as $g g \rightarrow n\cdot Z$ is that they only involve fermion loops of maximal numerator rank. Contrary to the stability studies performed on all other classes of processes, we investigated the loop-induced ones by considering only a single helicity configuration; either $g^{(-)} g^{(-)} \rightarrow n\cdot Z^{(-)}$ (all-minus) or $g^{(-)} g^{(-)} \rightarrow n\cdot Z^{(0)}$ (all longitudinal Z bosons).
We find a similar stability behavior for both helicity configurations, except for the lowest multiplicity process. 

The peculiarity of the $g g \rightarrow Z Z$ process is that its fermion box loop contribution becomes unstable when the transverse momentum $p_t$ of the final state $Z$-bosons tends to zero (i.e.\ all external momenta aligned on the beam axis). Given the constrained $2\rightarrow 2$ kinematics, this configuration is often probed which is also why this process is typically integrated using a technical (very) small cut in the $Z$-boson $p_t$. The upper left plot of fig.~\ref{fig:StabilityPlot_zz} reveals that in this $p_t \rightarrow 0$ limit, the fermion box stability significantly depends on the helicity configuration considered\footnote{The numerical stability of the process $g g \rightarrow Z Z$ is actually minimal for the helicity configuration '-+00' (and its parity-related counterpart '+-00'), not shown in fig.~\ref{fig:StabilityPlot_zz}.}. For more than two $Z$-bosons in the final states, this dependence is much weaker, and mainly reflects the difference in the size of the relative contribution of the more stable Higgs channels.

Even though the highest multiplicity loop-induced process is of maximal rank 7, it is significantly more stable than $g g \rightarrow Y g g g$ which shares the same maximal rank but with lower multiplicity. On the other hand, its stability is on par with the other $2\rightarrow5$ process $g g \rightarrow t \bar{t} g g g$ which has many more diagrams but with a maximal rank of only 6.\\

In summary and in terms of numerical stability, \Ninja\ performs better or at least as well as the best other public reduction tool available for all considered processes except two. First, contrary to \Golem, \Ninja\ has around 1\% of unstable kinematic configurations (featuring less than 3 digits) for $g g \rightarrow H g g$ and secondly, it is slightly less stable than \CutTools\ for $u \bar{u} > 5\cdot Z$.

The observations drawn in this appendix suggest that the numerical stability of one-loop matrix elements can be \emph{classified} according to mainly their maximal rank and multiplicity. However, a large difference in the number and complexity of the contributing loops (in terms of their tensor numerator structure and typical rank, as well as the number and spread of the different scales) can amount to variations as large as the gap between two consecutive such stability \emph{classes}.

\section{Numerical results for benchmark processes}
\label{sec:numresults}

In this appendix, we include numerical evaluations of the one-loop matrix elements of the processes $g g \rightarrow t \bar{t} g g g$ and $g g \rightarrow Y + \{2,3\}\cdot g$, for selected random kinematic configurations and summed over helicity and colour assignments.
Although the generation of these processes is completely automated within \MGaMC, their complexity makes them both memory and CPU intensive, so that we report here numerical results for convenience, and also because it is, to the best of our knowledge, not available in the literature. It can serve as a benchmark for validating less automated optimized implementations, potentially fast enough to render phase-space integration feasible.
We express the numerical value of the one-loop matrix element in terms of the three coefficients $c_{-2}$, $c_{-1}$ and $c_0$, implicitly defined as follows
\begin{equation}
\frac{1}{S_F}\overline{\mathop{\sum_{\rm colour}}_{\rm spin}}
2\Re\left\{\mathcal{A}^{(1)}\mathcal{A}^{(0)\star}\right\}=\frac{(4\pi)^\epsilon}{\Gamma(1-\epsilon)}
\left(\frac{c_{-2}}{\epsilon^2}+\frac{c_{-1}}{\epsilon}+c_0\right)\,,
\end{equation}
and similarly for the Born contribution
\begin{equation}
\frac{1}{S_F}\overline{\mathop{\sum_{\rm colour}}_{\rm spin}}
\left |\mathcal{A}^{(0)}\right |^2=a_0\,.
\end{equation}
The $\overline{\Sigma}$ symbol denotes that we sum over all final-state helicity and color configurations but we average over initial-state ones. We account for the conventional symmetry factor $S_F$ equal to the product of the factorial of the number of identical copies of any particle in the final states (so $S_F=n_{\rm{final\_gluons}}!$ for the processes of interest here).

We remind the reader that numerical results like the ones presented in this appendix, along with a library for the corresponding computation for any phase-space point\footnote{See \href{https://cp3.irmp.ucl.ac.be/projects/madgraph/wiki/MadLoopStandaloneLibrary}{cp3.irmp.ucl.ac.be/projects/madgraph/wiki/MadLoopStandaloneLibrary} for more information.}, can be obtained by running the following three commands in the interactive interface of \MGaMC:

\noindent\\
~\prompt\ {\tt ~import <chosen\_model>}\\
~\prompt\ {\tt ~generate <process\_definition> [virt=QCD]}\\
~\prompt\ {\tt ~launch}\\

The first step is optional when considering corrections for processes within the Standard Model. For mixed Electro-Weak (EW) and QCD correction, the syntax '{\tt [virt=QCD QED]}' must be used instead.
\noindent
 
\subsection{Process $g g \rightarrow t \bar{t} g g g$}
\label{sec:ggttxggg}

We specify below the chosen values for the relevant Standard Model parameters. Notice that all dimensionful quantities in this Appendix~\ref{sec:numresults} are indicated with GeV units, unless specified otherwise.
\begin{center}
\begin{tabular}{cl|cl}\toprule
Parameter & value & Parameter & value
\\\midrule
$\as$ & \texttt{0.118} & $n_{lf}$ & \texttt{4}
\\
$m_{t}$ & \texttt{173.0}    & $m_{b}$ & \texttt{4.7}
\\
$\Gamma_{t}$ & \texttt{0.0} & $\mu_r$ & 91.188
\\\bottomrule
\end{tabular}
\end{center}
The kinematic configuration considered is:
\begin{footnotesize}
  \begin{align*}
                      & \phantom{=\textrm{(--}}\textrm{E}  && \phantom{\textrm{,--}}p_x &&\phantom{\textrm{,--}}p_y &&\phantom{\textrm{,--}}p_z\\
    p_g            &=\textrm{( 500}&&                           \textrm{, \phantom{-}0}                               &&\textrm{, \phantom{-}0}                               &&\textrm{, \phantom{-}500}                           &&\textrm{)}\\    
    p_g            &=\textrm{( 500}&&                           \textrm{, \phantom{-}0}                               &&\textrm{, \phantom{-}0}                               &&\textrm{, -500}                                            &&\textrm{)}\\
    p_t             &=\textrm{( 187.1526174824760}&&\textrm{, -12.57780476763981}&&\textrm{, \phantom{-}31.11387284817540}&&\textrm{, -63.01450606135616}                  &&\textrm{)}\\   
    p_{\bar{t}}  &=\textrm{( 289.0726872654491}&&\textrm{, -68.72525125230017}&&\textrm{, -217.6061267288446}&&\textrm{, \phantom{-}39.47698029543776}                  &&\textrm{)}\\
    p_g             &=\textrm{( 105.4303841594690}&&\textrm{, -74.77954838939333}&&\textrm{, -68.76840480238191}&&\textrm{, \phantom{-}28.18672644397240}                  &&\textrm{)}\\   
    p_g  &=\textrm{( 342.3857129250602}&&\textrm{, \phantom{-}191.9284080684538}&&\textrm{, \phantom{-}275.1076866100379}&&\textrm{, -68.60920754231151}                  &&\textrm{)}\\
    p_g             &=\textrm{( 75.95859816754576}&&\textrm{, -35.84580365912052}&&\textrm{, -19.84702792698694}&&\textrm{, \phantom{-}63.96000686425754}                  &&\textrm{)}    
  \end{align*}
\end{footnotesize}
We report below all stable digits (no rounding applied) obtained with \emph{double precision arithmetics} for the coefficients $a_{0}$, $c_{-2}$, $c_{-1}$ and $c_{0}$.
\begin{center}
\begin{tabular}{cl}\toprule
  $[\rm{GeV}^{-6}]$ & \multicolumn{1}{c}{ $g g \to t\bar{t} g g g$}
  \\\midrule
$a_0$ & \texttt{\phantom{-}1.5516406229353492e-9}
\\
$c_{0}$ & \texttt{-5.6512364e-10}
\\
$c_{-1}$ & \texttt{\phantom{-}3.504298678e-10}
\\
$c_{-2}$ & \texttt{-4.371037568250e-10}
\\\bottomrule
\end{tabular}
\end{center}

\subsection{Processes $g g \rightarrow Y + \{2,3\}\cdot g$}
\label{sec:ggY23g}

The details of the model that we considered including the spin-2 particle $Y$ can be found in ref.~\cite{Artoisenet:2013puc}. 
As already noted in~\cite{Mathews:2004xp}, the operator renormalization constant for the energy momentum operator is
identical to unity to all orders in perturbation theory. As a result, when the graviton minimally couples to the energy momentum tensor (i.e.\ with $\kappa_q=\kappa_g$), there is no need for additional UV renormalisation counterterms.
The parameters of this model that are relevant for the processes $g g \rightarrow Y + \{2,3\}\cdot g$ are chosen as follows:
\begin{center}
\begin{tabular}{cl|cl}\toprule
Parameter & value & Parameter & value
\\\midrule
$\as$ & \texttt{0.118} & $n_{lf}$ & \texttt{5}
\\
$m_{t}$ & \texttt{173.0}    & $\mu_r$ & 91.188
\\
$\Gamma_{t}$ & \texttt{0.0} & $m_{Y}$ & \texttt{300}
\\
$\kappa_{g}=\kappa_{q}$ & \texttt{1.0} & $\Lambda$ & \texttt{1000}
\\\bottomrule
\end{tabular}
\end{center}
The $2\rightarrow3$ and $2\rightarrow4$ kinematic configurations considered are
\begin{footnotesize}
  \begin{align*}
             	& \phantom{=\textrm{(--}}\textrm{E}  && \phantom{\textrm{,--}}p_x &&\phantom{\textrm{,--}}p_y &&\phantom{\textrm{,--}}p_z\\
    p_g    	&=\textrm{( 500}&&                           \textrm{, \phantom{-}0}                               &&\textrm{, \phantom{-}0}                               &&\textrm{, \phantom{-}500}                           &&\textrm{)}\\    
    p_g   	&=\textrm{( 500}&&                           \textrm{, \phantom{-}0}                               &&\textrm{, \phantom{-}0}                               &&\textrm{, -500}                                            &&\textrm{)}\\
    p_Y	&=\textrm{( 511.2453080350917}&&\textrm{, \phantom{-}152.9697297441591}&&\textrm{, \phantom{-}342.7230121584469}&&\textrm{, -174.6796030077232}                  &&\textrm{)}\\   
    p_g	&=\textrm{( 328.6521936921786}&&\textrm{, -16.54683898542085}&&\textrm{, -313.8815119978393}&&\textrm{, \phantom{-}96.00449450011196}                  &&\textrm{)}\\   
    p_g    	&=\textrm{( 160.1024982727296}&&\textrm{, -136.4228907587383}&&\textrm{, -28.84150016060760}&&\textrm{,  \phantom{-}78.67510850761127}                  &&\textrm{)}
  \end{align*}
\end{footnotesize}
and
\begin{footnotesize}
  \begin{align*}
             	& \phantom{=\textrm{(--}}\textrm{E}  && \phantom{\textrm{,--}}p_x &&\phantom{\textrm{,--}}p_y &&\phantom{\textrm{,--}}p_z\\
    p_g    	&=\textrm{( 500}&&                           \textrm{, \phantom{-}0}                               &&\textrm{, \phantom{-}0}                               &&\textrm{, \phantom{-}500}                           &&\textrm{)}\\    
    p_g   	&=\textrm{( 500}&&                           \textrm{, \phantom{-}0}                               &&\textrm{, \phantom{-}0}                               &&\textrm{, -500}                                            &&\textrm{)}\\
    p_Y	&=\textrm{( 187.1526174824760}&&\textrm{, -12.57780476763981}&&\textrm{, \phantom{-}31.11387284817540}&&\textrm{, -63.01450606135616}                  &&\textrm{)}\\   
    p_g	&=\textrm{( 105.4303841594690}&&\textrm{, -74.77954838939333}&&\textrm{, -68.76840480238191}&&\textrm{, \phantom{-}28.18672644397240}                  &&\textrm{)}\\   
    p_g    	&=\textrm{( 342.3857129250602}&&\textrm{, \phantom{-}191.9284080684538}&&\textrm{, \phantom{-}275.1076866100379}&&\textrm{, -68.60920754231151}                  &&\textrm{)}\\
    p_g     	&=\textrm{( 75.95859816754576}&&\textrm{, -35.84580365912052}&&\textrm{, -19.84702792698694}&&\textrm{, \phantom{-}63.96000686425754}                  &&\textrm{)}    
  \end{align*}
\end{footnotesize}
We conclude by reporting below all stable digits obtained with \emph{quadruple precision arithmetics} for the coefficients $a_{0}$, $c_{-2}$, $c_{-1}$ and $c_{0}$
\begin{center}
\begin{tabular}{cll}\toprule
  $[\rm{GeV}^{-2,-4}]$ & \multicolumn{1}{c}{ $g g \to Y g g$} & \multicolumn{1}{c}{ $g g \to Y g g g$}
  \\\midrule
$a_0$ & \texttt{\phantom{-}1.864507929648605e-3}		&	\texttt{\phantom{-}1.529529540966328e-6}
\\
$c_{0}$ & \texttt{-1.003108324246923e-3}				&	\texttt{-4.0992246274118e-7}
\\
$c_{-1}$ & \texttt{\phantom{-}8.424377254497548e-4}	&	\texttt{\phantom{-}5.5831926923384e-7}
\\
$c_{-2}$ & \texttt{-4.201918452676584e-4}			&	\texttt{\phantom{-}4.308749710782677e-7}
\\\bottomrule
\end{tabular}
\end{center}

\end{appendices}

\bibliographystyle{JHEP}
\bibliography{biblio}

\end{document}